\documentclass[amsmath,amssymb,11pt]{article}
\usepackage{jheppub2}
\pdfoutput=1
\usepackage{epsfig}
\usepackage{float}
\usepackage{subfig}
\usepackage{subfloat}
\usepackage{graphicx}
\usepackage[utf8]{inputenc}
\usepackage{hyperref}
\usepackage{caption}

\usepackage{color}

\global\long\def\G{\Gamma}

\global\long\def\pd{\dot{\phi}}
\global\long\def\pdd{\ddot{\phi}}

\global\long\def\mR{\mathcal{R}}
\global\long\def\mRd{\dot{\mathcal{R}}}
\global\long\def\mRdd{\ddot{\mathcal{R}}}

\global\long\def\l{\lambda}
\global\long\def\p{\partial}
\global\long\def\g{\gamma}
\global\long\def\z{\zeta}
\global\long\def\s{\sigma}
\global\long\def\a{\alpha}
\global\long\def\d{\delta}
\global\long\def\gdo{g^{(0)}}
\global\long\def\gdi{g^{(1)}}
\global\long\def\gdii{g^{(2)}}
\global\long\def\guo{g_{(0)}}
\global\long\def\gui{g_{(1)}}
\global\long\def\guii{g_{(2)}}
\global\long\def\dK{\dot{K}}
\global\long\def\da{\dot{a}}
\global\long\def\dda{\ddot{a}}
\global\long\def\ds{\dot{\sigma}}
\global\long\def\dH{\dot{H}}
\global\long\def\dz{\dot{\zeta}}
\global\long\def\dal{\dot{\alpha}}
\global\long\def\ddz{\ddot{\zeta}}
\global\long\def\Go{[\Gamma^{(0)}]}
\global\long\def\Gi{[\Gamma^{(1)}]}
\global\long\def\Gii{[\Gamma^{(2)}]}

\global\long\def\bP{\bar{P}}
\global\long\def\bA{\bar{A}}
\global\long\def\bB{\bar{B}}
\global\long\def\bC{\bar{C}}
\global\long\def\bD{\bar{D}}
\global\long\def\ta{\tilde{a}}
\newcommand*{\affmark}[1][*]{\textsuperscript{#1}}

\newcommand{\beq}{\begin{equation}}

\newcommand{\eeq}{\end{equation}}

\title{Cosmic acceleration from string induced Galileons}

\author{Damien A. Easson,\affmark[1]}
\emailAdd{deasson@asu.edu}
\author{Tucker Manton,\affmark[1]}
\emailAdd{tucker.manton@asu.edu}
\author{and Andrew Svesko\affmark[2]}
\emailAdd{a.svesko@ucl.ac.uk}

\affiliation{\affmark[1]Department of Physics, Arizona State University, Tempe, Arizona 85287, USA\\
\affmark[2]Department of Physics and Astronomy, University College London,\\
Gower Street, London, WC1E 6BT, United Kingdom}

\abstract{It has been shown a specific Horndeski theory of gravity arises from a consistent Kaluza-Klein reduction of the gravi-dilaton sector of the low-energy effective heterotic string action with a first $\alpha'$ correction. Here we provide a first investigation of the cosmological solutions to the lower dimensional theory by constructing exact solutions to the field equations in various frames. At tree level, in string frame, we find the duality symmetry of the parent theory is unaltered by dimensional reduction, leading to standard bouncing and pre-Big Bang scenario solutions. For $\alpha'\neq0$, we uncover exact models exhibiting cosmic acceleration, where we find interesting physics concerning the number of noncompact dimensions. Certain stability requirements demand the inclusion of a dilaton potential.}

\begin{document}

\maketitle

\section{Introduction} \label{sec:intro}
\indent

Despite its experimental successes, general relativity (GR) provides an incomplete description of gravity. For one, GR alone is unable to account for the observed rotation curves of spiral galaxies, or the  current accelerated expansion of the universe. More fundamentally, general relativity predicts the existence of past and future spacetime singularities -- locations where Einstein's equations breakdown -- GR is not consistent with the omnipresent quantum world.  Motivated by these limitations, theorists have looked to modify general relativity so as to retain empirical low energy accuracy, while resolving its limitations. Indeed, the non-renormalizability of GR requires Einstein's theory to include higher curvature corrections, while certain candidate models of quantum gravity, such as string theory, tell us GR should include fields other than the metric. Higher derivative and non-minimally coupled theories of gravity have also emerged as potential alternatives to dark matter and dark energy. Thus, from both a fundamental and practical perspective, general relativity is to be replaced by a more complicated, potentially non-minimally coupled theory of spacetime.\footnote{For a broad overview of such modified theories of gravity, see, \emph{e.g.}, \cite{Clifton:2011jh}.}

Among the diverse alternatives to general relativity, two in particular emerge as elegant modifications: Lovelock \cite{Lovelock:1971yv,Lovelock:1972vz} and Horndeski gravity \cite{Horndeski74-1}.  Lovelock gravity is a pure diffeomorphism invariant theory of gravity with only second order equations of motion, thereby precluding the possibility of ghost-like Ostrogradsky instabilities. For spacetime dimensions $D=3,4$, general relativity is the unique theory satisfying Lovelock's theorem. For $D\geq5$, the Lovelock action includes higher curvature contributions, such that for dimension $D$, 
\beq I_{\text{ELL}}=\int d^{D}x\sqrt{-g}\left[\frac{1}{16\pi G_{D}}(R-2\Lambda_{0})+\mathcal{L}_{LL}\right]\;,\label{ELLactionD}\eeq
where $\Lambda_{0}$ is the bare cosmological constant, $G_{D}$ is the $D$-dimensional Newton's constant and the Lovelock Lagrangian is 
\beq \mathcal{L}_{\text{LL}}=\sum_{n=0}^{t}\mathcal{L}_{(n)}=\sum_{n=0}^{t}\alpha_{n}\mathcal{R}_{(n)}\;,\quad \mathcal{R}_{(n)}=\frac{1}{2^{n}}\delta^{\mu_{1}\nu_{1}...\mu_{n}\nu_{n}}_{\alpha_{1}\beta_{1}...\alpha_{n}\beta_{n}}\prod_{r=1}^{n}R^{\alpha_{r}\beta_{r}}_{\;\;\;\;\;\;\;\mu_{r}\nu_{r}}\;,\eeq
with $\alpha_{n}$ being Lovelock coupling constants and where $\delta^{\mu_{1}\nu_{1}...\mu_{n}\nu_{n}}_{\alpha_{1}\beta_{1}...\alpha_{n}\beta_{n}}$ denotes the generalized Kronecker delta symbol. For example, in $D=5$ Einstein gravity may be generalized to Einstein-Gauss-Bonnet (EGB) gravity,\footnote{Recently, pure Einstein-Gauss-Bonnet gravity was shown to exist in $D\leq4$ upon a clever dimensionful rescaling of the Gauss-Bonnet coupling constant \cite{Glavan:2019inb}, thereby evading Lovelock's theorem.} with Lagrangian, 
\beq \mathcal{L}_{\text{EGB}}=R+\alpha_{\text{GB}}(R^{2}-4R_{\mu\nu}^{2}+R_{\mu\nu\rho\sigma}^{2})\;.\eeq
As general relativity is a special case of $\mathcal{L}_{\text{LL}}$, Lovelock gravity is often viewed as the most natural modification of GR, provided one allows for $D\geq4$.

By Lovelock's theorem, any $D\leq4$ dimensional modification of gravity thus involves the addition of a non-trivial field, the simplest being a scalar $\phi$.\footnote{$f(R)$ theories of gravity seemingly evade this, however, it is well known $f(R)$ gravity is equivalent to the Brans-Dicke theories of gravity, at the classical and quantum level.} Horndeski gravity, or, equivalently, generalized covariantized Galileons \cite{Deffayet:2009wt,Deffayet:2009mn,Deffayet:2011gz}, is a preferred  non-minimally coupled scalar-tensor theory of gravity as it only has second order equations of motion. The covariant \emph{generalized} Galileons Lagrangian (as written in, \emph{e.g.}, \cite{Kobayashi:2011nu,Charmousis14-1}) is 
 \beq 
 \begin{split}
 \mathcal{L}&=K(\phi,X)-G_{3}(\phi,X)\Box\phi+G_{4}(\phi,X)R+G_{4,X}\left[(\Box\phi)^{2}-(\nabla_{\mu}\nabla_{\nu}\phi)^{2}\right]\\
 &+G_{5}(\phi,X)G_{\mu\nu}\nabla^{\mu}\nabla^{\nu}\phi-\frac{1}{6}G_{5,X}\left[(\Box\phi)^{3}-3\Box\phi(\nabla_{\mu}\nabla_{\nu}\phi)^{2}+2(\nabla_{\mu}\nabla_{\nu}\phi)^{3}\right]\;,
 \end{split}
\label{Horndaction} \eeq
where $X\equiv-\frac{1}{2}(\nabla\phi)^{2}$, and $K,$ $G_{3,4,5}$ are general functions of $\phi$ and $X$, with $G_{i,X}\equiv\frac{\partial G_{i}}{\partial X}$ and similarly for $G_{i,\phi}$. Originally a local modification of general relativity \cite{Nicolis:2008in}, Galileons are particularly relevant in cosmology, where they have been used to describe the inflationary epoch of the universe \cite{Silva:2009km,Creminelli10-1,Kobayashi:2010cm}, dark energy \cite{Deffayet:2010qz}, and bouncing cosmologies \cite{Qiu:2011cy,Easson:2011zy,Cai:2012va,Rubakov13-1,Ijjas16-1}. For a review on the various scalar-tensor theories of gravity embedded in (\ref{Horndaction}) and their current status in light of the neutron star merger event GW170817\footnote{The announcement of this observation can be found in \cite{Monitor:2017mdv}, where a bound on the difference between the speed of light and the speed of the gravitational waves is provided. This is relevant to some of our results and will be discussed further in section \ref{sec:stability}. }, see \cite{Kase:2018aps}.


While both of these alternatives to general relativity are well motivated by symmetry principles and for their practical advantages, both lack a fundamental origin. That is, neither Horndeski or Lovelock gravity represent complete theories, and generally continue to suffer from the same non-renormalizability issues that plague GR (look up trodden). The Gauss-Bonnet term, however, is believed to arise as the first $\alpha'$ correction to the low energy effective action of 10-dimensional heterotic string theory \cite{Zwiebach85-1,Sen:1985qt,Gross86-1,Gross:1986mw,Metsaev:1987zx}. In this way, the Gauss-Bonnet term can naturally appear in a UV complete model of quantum gravity. Separately, the authors of \cite{VanAcoleyen11-1} demonstrated a specific class of Horndeski theories of gravity emerge from higher-dimensional Lovelock theories of gravity via Kaluza-Klein (KK) dimensional reduction. A consistent dimensional reduction of a pure higher derivative theory of gravity with second order equations of motion is guaranteed to yield a non-minimally coupled theory of gravity with second order field equations.\footnote{Combining the observation of \cite{VanAcoleyen11-1} together with the dimensionful rescaling utilized in \cite{Glavan:2019inb}, an alternative `novel' theory of $D=4$ Einstein-Gauss-Bonnet gravity was discovered in \cite{Lu:2020iav} (shown to be equivalent to dimensionally regularized EGB in \cite{Fernandes:2020nbq,Hennigar:2020lsl,Easson:2020mpq}), circumventing some of the ill-posedness of the pure theory. The authors of \cite{Easson:2020mpq} adapted the method of dimensional regularization to generic Lovelock theories of gravity, and $\alpha'$-corrected heterotic string theory.}

Combining these two observations -- the Gauss-Bonnet term appears in string theory, and Horndeski gravity emerges from a consistent KK reduction of the Lovelock action -- the authors of \cite{Easson:2020bgk} demonstrated Galileons may have a string theoretic origin by dimensionally reducing the first $\alpha'$-corrected heterotic string action. In this article, we begin the first of multiple investigations into the types of spacetimes governed by the field equations of this string induced Horndeski model of \cite{Easson:2020bgk}, beginning with cosmological solutions. In particular we find exact solutions to the background equations of motion in an FLRW ansatz supported by the dilaton, which acts as the Horndeski scalar in arbitrary frames (add footnote). Power-law type solutions and constant curvature solutions are found in specific frames of interest, \emph{e.g.}, Einstein frame, which also admit cosmic acceleration. Specializing the $3+1$-dimensions, we find the exact solutions are unstable against scalar or tensor perturbations, however, stabilization occurs for various dilaton potentials, found by numerically solving the field equations.

This article is organized as follows. In Section \ref{sec:review} we briefly review how the generalized Galileons emerge from KK reduction of pure Einstein-Gauss-Bonnet gravity and then from an $\alpha'$-corrected effective string action. We point out additonal features the string induced Galileons exhibit (previously unstated in \cite{Easson:2020bgk}), including a brief investigation into the symmetry properties of the tree-level action, as well as a discussion about the string coupling upon KK reduction. Section \ref{sec:FRLWcosmo} considers an FLRW metric ansatz where we find exact solutions to the field equations in arbitrary frames. We then specialize to specific frames, uncovering logarithmic and linear dilaton solutions that drive cosmic acceleration. In Section \ref{sec:stability}, we restrict to $3+1$-dimensions and provide a stability analysis of the exact solutions by studying the propagation speed of tensorial and scalar perturbations.  Each frame is unstable in some way, leading us to introduce a dilaton potential for which, specifically, we show accelerating solutions in Einstein frame become stabilized. We conclude in Section \ref{sec:disc} with a brief summary of the main results and point out how the model of string induced Galileons evades the no-go-like criteria presented in \cite{Montefalcone:2020vlu}. For completeness, we also include a number of appendices. In Appendix \ref{sec:apploweeffact} we provide a lightning review of the low energy effective string actions of interest to string cosmology. Appendix \ref{AppendixForRates} collects the numerical data used in our analysis of exact cosmological solutions. Stability criterion of tensorial and scalar perturbations about FLRW spacetimes for general Horndeski theories is summarized in Appendix \ref{app:linperts}.

\section{Galileons from dimensionally reduced string theory} \label{sec:review}

\subsection{Galileons from Lovelock: review}

Generalized Galileons can arise from Kaluza-Klein reduction of the Lovelock action \cite{VanAcoleyen11-1}. In particular, consider pure $D$-dimensional Einstein-Gauss-Bonnet gravity, defined by
\beq I_{\text{EGB}}=\int \hspace{-1mm}d^{D}x\sqrt{-\hat{g}}\left[\frac{1}{16\pi G_{D}}(\hat{R}-2\Lambda_{0})+\hat{\mathcal{L}}_{\text{GB}}\right]\,. \label{EGBactionD}\eeq
 Now, decompose the $D$-dimensional line element $d\hat{s}^{2}=\hat{g}_{MN}dX^{M}dX^{N}$ as
\beq d\hat{s}^{2}_{(p+1+n)}=e^{2\alpha\phi}ds^{2}_{p+1}+e^{2\beta\phi}dK^{2}_{(n)}\;, \label{metansatz2}\eeq
where $\alpha$ and $\beta$ are (mostly) unspecified real constants -- except that for consistency we cannot have $\alpha=\beta=0$ --  and $D=p+1+n$. The (uncompactified) $p+1$-dimensional geometry is characterized by the line element $ds^{2}_{p+1}=g_{\mu\nu}dx^{\mu}dx^{\nu}$ and is described by coordinates $\{x^{\mu}\}$, with $\mu=0,...,p$, while $n$-dimensional internal space we compactify over is generically described by the metric $dK_{(n)}^{2}=\tilde{g}_{ij}(dz^{i}+A_{\mu}^{(i)}dx^{\mu})(dz^{j}+A_{\nu}^{(j)}dx^{\nu})$ with coordinates $\{z^{i}\}$, $i=1,...,n$. Here $A^{(i)}_{\mu}$ is a vector potential associated with each curled up dimension, and can be interpreted as a gauge field of some Maxwell- or Yang-Mills-like gauge theory from the $p+1$-dimensional perspective. The field $\phi$ is present to maintain the proper number of degrees of freedom of the higher dimensional metric, however, as we will see shortly, is interpreted as the scalar present in single-field Horndeski gravity, or the dilaton in string theory.

 For simplicity we will assume $dK^{2}_{(n)}$ is flat, and, moreover, we consider a \emph{diagonal} dimensional reduction, \emph{i.e.}, each vector potential $A^{(i)}_{\mu}=0$. We further assume the $p+1$-dimensional metric $g_{\mu\nu}$ and scalar field $\phi$ are only functions of the $p+1$-dimensional coordinate system, \emph{ i.e.,} $g_{\mu\nu}=g_{\mu\nu}(x)$ and $\phi=\phi(x)$. This amounts to supplementing the metric ansatz (\ref{metansatz2}) with the standard assumption of keeping only the lowest modes of the KK tower,\emph{ i.e.,} the lower dimensional metric $g_{\mu\nu}=g_{\mu\nu}(x)$ and $\phi=\phi(x)$. This assumption is justified since the isometry group of the compactified internal space  is Abelian, in which case the massive modes of the KK reduction are truncated away \cite{Gouteraux:2011qh}. 

Crucially, the ansatz (\ref{metansatz2}), including the simplifying assumptions described above, was explicitly shown to allow for consistent dimensional reductions of EGB gravity in \cite{Gouteraux:2011qh}, such that  all higher dimensional solutions to the pure EGB theory respecting the symmetries of the ansatz (\ref{metansatz2}) will be solutions to the reduced theory, and, vice versa, all lower dimensional solutions can be oxidized to $p+1+n$ dimensions, likewise respecting the symmetries of the solutions to the higher dimensional theory. Therefore, substituting the ansatz (\ref{metansatz2}) into the action (\ref{EGBactionD}), one finds \cite{Gouteraux:2011qh,Charmousis12-1} (see Appendix A of \cite{Easson:2020bgk} for further details)
\beq
\begin{split}
\sqrt{-\hat{g}}\hat{R}&=\sqrt{-g}e^{[(p-1)\alpha+n\beta]\phi}\biggr\{R+Y\Box\phi+Z(\nabla\phi)^{2}\biggr\}\;,
\end{split}
\label{EHterm1}\eeq
\beq \begin{split}
\sqrt{\hat{g}}\hat{\mathcal{L}}_{GB}&=\sqrt{-g}e^{[(p-3)\alpha+\beta n]\phi}\biggr\{\mathcal{L}_{GB}+B_{1}G^{\mu\nu}(\nabla_{\mu}\phi)(\nabla_{\nu}\phi)+B_{2}(\nabla\phi)^{4}+B_{3}\Box\phi(\nabla\phi)^{2}\biggr\}\;.
\end{split}
\label{Ghatscaltenhighv3}\eeq
where the (unique) coefficients $Y,Z$, and $\{B_{i}\}$ are complicated polynomials of $p,n,\alpha$ and $\beta$. We present the full expressions for the coefficients in the context of the reduced string model momentarily.

Focusing on the reduced Gauss-Bonnet action (\ref{Ghatscaltenhighv3}), we observe the emergence of Horndeski gravity, where the coefficients and functions of $X$ and $\phi$ take a particular form. Specifically, in (\ref{Ghatscaltenhighv3}) we have exchanged $G^{\mu\nu}\nabla_{\mu}\nabla_{\nu}\phi$ for a $G^{\mu\nu}(\nabla_{\mu}\phi)(\nabla_{\nu}\phi)$ term, up to total derivatives, and similarly for $[(\nabla_{\mu}\nabla_{\nu}\phi)^{2}-(\Box\phi)^{2}]$. Note that the reduction of the Gauss-Bonnet term alone does not yield the six-derivative quantity $G_{5,X}$ in the Galileon action (\ref{Horndaction}). Such a term may be generated by KK reduction of Lovelock terms beyond Gauss-Bonnet \cite{VanAcoleyen11-1,Charmousis:2011bf}. Moreover, due to the presence of the conformal factor, the KK reduced Horndeski theory lacks a constant shift symmetry in $\phi$.\footnote{When not coupled to gravity, Galileon theories have the interpretation of a Goldstone boson $\phi$ in flat space which is invariant under the shift symmetry $\phi(x)\to\phi(x)+b_{\mu}x^{\mu}+c$ for arbitrary parameters $b^{\mu}$ and $c$. As modifications to general relativity \cite{Nicolis:2008in}, some Horndeski theories preserve the constant shift symmetry.}


\subsection{String induced Galileons}

Building off of the observations made in \cite{VanAcoleyen11-1}, the authors of \cite{Easson:2020bgk} considered the first $\alpha'$ correction to the low-energy effective action of 10-dimensional heterotic string theory, which famously includes a Gauss-Bonnet term \cite{Zwiebach85-1,Sen:1985qt,Gross86-1,Gross:1986mw,Metsaev:1987zx}:
\beq I=-\frac{1}{2\lambda^{D-2}_{s}}\int d^{D}X\sqrt{-\hat{g}}e^{-\Phi}\left[\hat{R}+(\hat{\nabla}\Phi)^{2}-\frac{\alpha'}{4}\hat{\mathcal{L}}_{GB}+\frac{\alpha'}{4}(\hat{\nabla}\Phi)^{4}\right]\;,\label{stringact1}\eeq
where $\Phi$ is the dilaton, $\lambda_{s}$ is the natural length scale associated with a 1-dimensional string,\footnote{Comparing to the Einstein-Hilbert action, the length scale $\lambda_{s}$ is related to the standard gravitational coupling via $8\pi G_{D}=\lambda^{D-2}_{s}\exp(\Phi)$.} and $\alpha'$ is inversely proportional to the string tension $T_{s}\equiv (2\pi\alpha')^{-1}$, and defines the fundamental string length $\alpha'\equiv\ell_{s}^{2}$. The Gauss-Bonnet term appears in order to rid the low-energy theory of any ghosts and for the $\alpha'$ corrected Polyakov string action to obey conformal invariance at the quantum level. We emphasize that here our focus is on the gravi-dilaton sector. More generally the string action will include the antisymmetric Kalb-Ramond field $B_{\mu\nu}$, contributions which may lead to higher derivative scalar field contributions that could be interpreted as multi-Galileons at the $p+1$-dimensional level \cite{Bakhmatov:2019dow}. For simplicity we focus on single field Galileons and therefore neglect these types of contributions.

Upon dimensional reduction, we arrive to the lower-dimensional action \cite{Easson:2020bgk}
\beq 
\begin{split}
 I&=-\frac{1}{2\lambda^{p-1}_{s}}\int d^{p+1}x\sqrt{-g}\biggr\{e^{[(p-1)\alpha+n\beta-1]\phi}\left[R+a_{1}(\nabla\phi)^{2}\right]\\
&-\frac{\alpha'}{4}e^{[(p-3)\alpha+n\beta-1]\phi}\left[\mathcal{L}_{GB}+a_{2}G^{\mu\nu}(\nabla_{\mu}\phi)(\nabla_{\nu}\phi)+a_{3}(\nabla\phi)^{2}\Box\phi+a_{4}(\nabla\phi)^{4}\right]\biggr\}\;,
\end{split} 
\label{redstringact}\eeq
where $\Phi(X^{N})=\phi(x^{\mu})$, we have integrated out the internal space into the adjusted parameter $\lambda^{p}_{s}$, and dropped total derivatives. Comparing to the standard Einstein-Hilbert action, we can read off the effective gravitational coupling in $p+1$ dimensions: $8\pi G_{p+1}=\lambda^{p-1}_{s}e^{[(p-1)\alpha+n\beta-1]\phi}$. We will have more to say about setting $\Phi(X^{N})=\phi(x^{\mu})$ momentarily. The coefficients $a_{i}$ are 
\beq\label{ais}
\begin{split}
&a_{1}\equiv 1+p^{2}\alpha^{2}+n\beta(-2+(n-1)\beta)-p\alpha(2+\alpha-2n\beta) \;,\\
&a_{2}\equiv -4(p-3)(p-2)\alpha^{2}+4n\beta(2-(n-1)\beta)-8\alpha(p-2)(n\beta-1)\;,\\
&a_{3}\equiv-2(p-1)(p-2)(p-3)\alpha^{3}-2n(n-1)\beta^{2}[-3+(n-2)\beta]\\
&-6n(p-1)\alpha\beta[-2+(n-1)\beta]-6(p-1)(p-2)\alpha^{2}(n\beta-1)\;,\\
&a_{4}\equiv-1-(p-1)(p-2)[2+(p-2)\alpha(-4+(p-3)\alpha)]\alpha^{2}-4n(n-1)^{2}[-1+(p-1)\alpha]\beta^{3}\\
&-4n(p-1)\alpha\beta[1+\alpha(5-3p+(p-2)^{2}\alpha)]-n(n-1)^{2}(n-2)\beta^{4}\\
&-2n[-1+n-2(3n-2)(p-1)\alpha+(p-1)(3-2p+n(3p-5))\alpha^{2}]\beta^{2}\;.
\end{split}
\eeq
The reduced action (\ref{redstringact}) is the Horndeski theory embedded in the gravi-dilaton sector of the higher dimensional superstring low energy effective action. In fact, comparing to the Horndeski Lagrangian (\ref{Horndaction}) for specific choices of $\alpha$ and $\beta$ one recovers special dilaton theories of gravity including Brans-Dicke, K-essence, and the `Fab Four'  \cite{Easson:2020bgk}. Particularly, the Horndeski functions as in the notation of (\ref{Horndaction}) are (in addition to the Gauss-Bonnet contribution)
    \begin{equation}
    \begin{split}
        K(\phi,X)&=-2a_1e^{[(p-1)\a+n\beta-1]\phi}X-4\frac{\alpha'}{4}a_4e^{[(p-3)\a+n\beta-1]\phi}X^2, \\
        G_3(\phi,X)&= -2\frac{\alpha'}{4}a_3e^{[(p-3)\a+n\beta-1]\phi}X, \\
        G_4(\phi)&=e^{[(p-1)\a+n\beta-1]\phi}, \\
        G_5(\phi) &=\frac{\alpha'}{4}\frac{a_2}{(p-3)\a+n\beta-1}e^{[(p-3)\a+n\beta-1]\phi}\;,
    \end{split}
\label{eq:formfactshorn}\end{equation}
We should emphasize that, due to the KK reduction mechanism, the functions $\{K,G_{3},G_{4},G_{5}\}$ take a specific form and are not completely arbitrary. This helps eliminate the otherwise potentially vast freedom in choosing the Horndeski functions, \emph{e.g.}, $G_{5,X}=0$. A caveat to this is that the string model (\ref{redstringact}) includes a Gauss-Bonnet term, which in four spacetime dimensions may be rewritten in terms of $\{K,G_{3},G_{4},G_{5}\}$ including a $G_{5,X}\neq0$ generally \cite{Kobayashi:2011nu,DeFelice:2011uc}. We will return to this point in Section \ref{sec:stability}.

Several more comments are in order. First and foremost, unlike the dimensional reduction of pure Einstein-Gauss-Bonnet gravity, the coefficients $\{a_{i}\}$ are not necessarily uniquely fixed. This is because the action (\ref{stringact1}) comes with an inherent field redefinition ambiguity. Indeed, the form of (\ref{stringact1}) arises from a specific choice in field redefinition, particularly the choice given by \cite{Metsaev:1987zx,Gasperini07-1}. Other field redefinitions allow one to write down an action which already includes many of the Galileon terms at the level of $D$-dimensional action. (see Appendix \ref{sec:apploweeffact} and similar discussion in \cite{Maeda:2011zn}). Dimensionally reducing such an action will also result in the emergence of Galileons in lower dimensions. Despite such field redefinitions, the ambiguity only results in constant shifts of the coefficients $a_{i}$, which are still  solely determined by $p,n,\alpha$ and $\beta$.

Secondly, different choices of $\alpha$ and $\beta$ allow us to place our theory in a particular `frame'. For example, when $\alpha=0$ and $\beta=\frac{1}{2}$, our Horndeski theory is said to be in the `dual frame', a frame which is of interest to studies in holography \cite{Kanitscheider:2009as,Gouteraux:2011qh}. The `Gauss-Bonnet frame' is one where we remove the conformal factor in front of the Gauss-Bonnet term, \emph{i.e.,} $[(p-3)\alpha+n\beta-1]=0$. In this frame, for $p\leq3$, the Gauss-Bonnet term is a total derivative and drops out from the dynamics of the theory.  Of particular interest to studies in cosmology are the `Einstein frame' and `string frame'. These frames amount to different choices of $\alpha$ and $\beta$. It is worth clarifying our frames are found by different choices of constants $\alpha$ and $\beta$: we are not performing field redefinitions or performing a conformal transformation. We will come back to this point later on. 

An additional relevant observation of the string induced Horndeski action (\ref{redstringact}), as with generic Horndeski theories, is that it may allow for the existence of stable bouncing cosmologies \cite{Qiu:2011cy,Easson:2011zy}. Such solutions require the Hubble parameter $H$ to satisfy $\dot{H}\geq0$, which, for ordinary general relativity is equivalent to violating the null energy condition (NEC). For those who prefer the NEC to be  understood solely as a criterion on the matter living on a spacetime, it is unclear whether the NEC is truly violated in the string induced model (\ref{redstringact}). This is because the $D$-dimensional heterotic string action includes non-minimal couplings between the metric and dilaton, thereby making it ambiguous to delineate the energy-momentum tensor purely for matter. Consequently, unlike the tree level ($\alpha'=0$) string action,\footnote{This is not entirely true, as there is already confusion how one should interpret energy conditions in Einstein frame versus string frame. This question was remedied in \cite{Chatterjee:2012zh} by modifying the string frame NEC such that it maps to the usual NEC obeyed in Einstein frame. Incidentally, both the modified string frame NEC and Einstein frame NEC obey the second law of thermodynamics, suggesting the NEC emerges from spacetime thermodynamics, as shown explicitly in \cite{Parikh:2015ret,Parikh:2016lys}.} it is unclear whether violations of the NEC in the parent theory map to NEC violations in the dimensionally reduced theory.

Let us now return to imposing $\Phi(X)=\phi(x)$. First note this arises from a consistent truncation of the higher modes in the Kaluza-Klein tower. Particularly, $\Phi(x,\{z_{i}\})=\sum_{k_{i}=0}\Phi_{k}(x)e^{2\pi i kz_{i}}\approx\Phi_{0}(x)\equiv\phi(x)$, where we separated $X^{N}=\{x^{\mu},z^{i}\}$ and dropped all of the modes from the $z_{i}$ directions. We could of course keep $\Phi$ and $\phi$ separate. That is, we can imagine integrating out the $\Phi(z^{i})$ modes, leaving a theory of two interacting scalar fields $\Phi(x^{\mu})$ and $\phi(x^{\mu})$, where the dilaton $\Phi$ is kept distinct from the Galileon $\phi$. Consequently, the reduced action is now an example of a multi-Galileon model. Assuming $\Phi(x)=\phi(x)$ leads to an important implication from the string theory perspective. Recall that the dilaton $\Phi$ controls the string coupling $g_{s}$;
\beq g_{s}^{2}=e^{\Phi}\;,\eeq
where an expansion in $g_{s}$ is equivalent to a genus expansion of the worldsheet geometries. Importantly, the conformal factor $e^{-\Phi}$ in the $D$-dimensional string action (\ref{stringact1}) is of the same order at both tree and $\alpha'$ level. That is, the low-energy effective action expanded to first order in $\alpha'$ is kept at fixed $g_{s}^{-2}$. Notably, however, in the reduced action (\ref{redstringact}) one observes the tree level and $\alpha'$ contributions are no longer at the same order in $g_{s}$ when $\Phi=\phi$ for arbitrary ansatz coefficients $\alpha$ and $\beta$. The dimensional reduction introduces a relative factor of $e^{-2\alpha\phi}=g^{-4\alpha}$ between the tree-level and $\alpha'$ contributions. We can recover the feature that both tree level and $\alpha'$ contributions are at the same order in a $g_{s}$ expansion if we set $\alpha=0$, $\beta\neq0$ -- the so-called generalized `dual frame'. This observation will be important for us later on when we attempt to understand the coupling regimes of our cosmological solutions.

Lastly, while we often consider the case where we have no dilaton potential, we can supplement the $D$-dimensional low energy effective string action (\ref{stringact1}) with 
\beq I^{D}_{V}=-\frac{1}{2\lambda_{s}^{D-2}}\int d^{D}X\sqrt{-\hat{g}}e^{-\Phi}(2\lambda^{D-2}_{s}V(\Phi))\,\eeq
which under KK reduction becomes 
\beq I_{V}^{p+1}=-\frac{1}{2\lambda_{s}^{p-1}}\int d^{p+1}x\sqrt{-g}e^{[(p-1)\alpha+n\beta-1]\phi}(2\lambda^{p-1}_{s}\tilde{V}(\phi))\;,\eeq
where we have identified $2\lambda^{p-1}_{s}\tilde{V}(\phi)\equiv 2\lambda^{D-2}_{s}e^{2\alpha\phi}V(\phi)$. The action $I_{V}^{p+1}$ is then added to the reduced string action (\ref{redstringact}).

Since they will be useful momentarily, for a Lagrangian of the generic form (where we explicitly include a dilaton potential $V(\phi)$) 
\beq \mathcal{L}=e^{A\phi}(R+b_{1}(\nabla\phi)^{2}-2\lambda_{s}^{p-1}V(\phi))-\frac{\alpha'}{4}e^{B\phi}\left[\mathcal{L}_{GB}+b_{2}G^{\mu\nu}(\nabla_{\mu}\phi)(\nabla_{\nu}\phi)+b_{3}\Box\phi(\nabla\phi)^{2}+b_{4}(\nabla\phi)^{4}\right]\;.\label{actiongen}\eeq
where, $A,B,b_{1},b_{2},b_{3}$, and $b_{4}$ are real coefficients and we have dropped the $\tilde{.}$ on $V(\phi)$, the dilaton and gravitational equations of motion are, respectively \cite{Easson:2020bgk}
\beq
\begin{split}
\mathcal{E}_{\phi}&=e^{A\phi}\left[AR-b_{1}A(\nabla\phi)^{2}-2b_{1}\Box\phi-2\lambda_{s}^{p-1}\left(AV(\phi)+\frac{dV}{d\phi}\right)\right]\\
&-\frac{\alpha'}{4}e^{B\phi}\biggr\{B\mathcal{L}_{GB}+(2R^{\mu\nu}-b_{2}BG^{\mu\nu})(\nabla_{\mu}\phi)(\nabla_{\nu}\phi)-2b_{2}G^{\mu\nu}(\nabla_{\mu}\nabla_{\nu}\phi)\\
&+(b_{3}B^{2}-3b_{4}B)(\nabla\phi)^{4}+(6b_{3}B-8b_{4})(\nabla\nabla\phi)(\nabla\phi)(\nabla\phi)-2b_{3}(\Box\phi)^{2}-4b_{4}\Box\phi(\nabla\phi)^{2}\biggr\}\;,
\end{split}
\label{phieom}\eeq
and
\beq
\begin{split}
\mathcal{E}_{\mu\nu}&= e^{A\phi}\biggr(G_{\mu\nu}-b_{1}[A\nabla_{\mu}\nabla_{\nu}\phi+A^{2}(\nabla_{\mu}\phi)(\nabla_{\nu}\phi)]+g_{\mu\nu}[A\Box\phi+A^{2}(\nabla\phi)^{2}]\\
&+b_{1}(\nabla_{\mu}\phi)(\nabla_{\nu}\phi)-\frac{b_{1}}{2}g_{\mu\nu}(\nabla\phi)^{2}+\frac{1}{2}g_{\mu\nu}2\lambda^{p-1}_{s}V(\phi)\biggr)\\
&+\frac{\alpha'}{8}e^{B\phi}\biggr\{g_{\mu\nu}b_{4}(\nabla\phi)^{4}+g_{\mu\nu}\mathcal{L}_{GB}+4R[B(\nabla_{\mu}\nabla_{\nu}\phi-g_{\mu\nu}\Box\phi)\\
&+B^{2}(\nabla_{\mu}\phi\nabla_{\nu}\phi-g_{\mu\nu}(\nabla\phi)^{2})]-[4RR_{\mu\nu}-8R_{\mu\gamma}R^{\gamma}_{\;\nu}-8R_{\mu\gamma\nu\delta}R^{\gamma\delta}+4R_{\mu\gamma\delta\rho}R_{\nu}^{\;\gamma\delta\rho}]\\
&+b_{2}\biggr[G_{\gamma\delta}(\nabla^{\gamma}\phi)(\nabla^{\delta}\phi)+2R_{\mu\delta\gamma\nu}(\nabla^{\gamma}\phi)(\nabla^{\delta}\phi)+g_{\mu\nu}R_{\gamma\delta}(\nabla^{\gamma}\phi)(\nabla^{\delta}\phi)\\
&-R_{\nu\delta}(\nabla_{\mu}\phi)(\nabla^{\delta}\phi)+R_{\mu\nu}(\nabla\phi)^{2}+2(\nabla_{\mu}\nabla_{\nu}\phi)\Box\phi-2(\nabla_{\mu}\nabla^{\gamma}\phi)(\nabla_{\nu}\nabla_{\gamma}\phi)+g_{\mu\nu}[(\nabla\nabla\phi)^{2}-(\Box\phi)^{2}]\\
&+B\biggr((\nabla_{\gamma}\phi)(\nabla_{\mu}\phi)(\nabla_{\nu}\nabla^{\gamma}\phi)+(\nabla_{\gamma}\phi)(\nabla_{\nu}\phi)(\nabla_{\mu}\nabla^{\gamma}\phi)-\Box\phi(\nabla_{\mu}\phi)(\nabla_{\nu}\phi)\\
&-g_{\mu\nu}(\nabla\nabla\phi)(\nabla\phi)(\nabla\phi)+(\nabla\phi)^{2}[g_{\mu\nu}\Box\phi-\nabla_{\mu}\nabla_{\nu}\phi]\biggr)\biggr]-4b_{4}(\nabla\phi)^{2}(\nabla_{\mu}\phi)(\nabla_{\nu}\phi)\\
&-b_{3}\biggr[g_{\mu\nu}[2(\nabla_{\alpha}\phi)(\nabla_{\beta}\phi)(\nabla^{\alpha}\nabla^{\beta}\phi)+B(\nabla\phi)^{4}]+2\Box\phi(\nabla_{\mu}\phi)(\nabla_{\nu}\phi)\\
&-2(\nabla^{\alpha}\phi)(\nabla_{\nu}\phi)(\nabla_{\alpha}\nabla_{\mu}\phi)-2(\nabla^{\alpha}\phi)(\nabla_{\mu}\phi)(\nabla_{\alpha}\nabla_{\nu}\phi)-2B(\nabla\phi)^{2}(\nabla_{\mu}\phi)(\nabla_{\nu}\phi)\biggr]\biggr\}\;.
\end{split}
\label{graveom}\eeq
In the following sections we will uncover exact cosmological solutions to these field equations in various frames of interest.


\subsection{String frame cosmology at tree level: duality and solutions} \label{sec:SFcosmo}

\subsection*{Tree level: duality and solutions}

Before we move on to consider cosmological solutions to the full $p+1$-dimensional theory in any frame, recall the $D$-dimensional string action (\ref{stringact1}) is of particular  interest in cosmological settings since it admits, for example, the `pre-big bang' scenario, bouncing cosmologies, and string gases (for reviews, see \cite{Veneziano:2000pz,Gasperini:2002bn,Gasperini07-1}). These types of solutions arise from the duality symmetry properties of the string action. It is natural to ask whether the dimensional reduction  breaks these symmetries, and if not, how might the interesting solutions to (\ref{stringact1}) appear in the reduced action (\ref{redstringact}). 

To answer this question, let us first focus on the reduced model (\ref{redstringact}) at tree level. Given that $\alpha$ and $\beta$ are mostly unspecified real parameters (recall for a consistent dimensional reduction $\alpha=\beta=0$ is not allowed), we have the freedom to choose $\alpha$ and $\beta$ such that the tree-level contribution (\ref{redstringact}) in $p+1$ dimensions is identical to the tree level contribution of the $D$-dimensional action (\ref{stringact1}). Specifically, demanding $(p-1)\alpha+n\beta=0$ and $a_{1}=1$ we find
\beq \alpha=-\frac{2n}{(p-1)(p+n-1)}\;,\quad \beta=\frac{(1-p)\alpha}{n}=\frac{2}{(p+n-1)}\;.\label{eq:alphbetsf}\eeq
In a slight abuse of terminology, when we choose $(\alpha,\beta)$ to take the values above, we say we are in `string frame'.

Since our choice (\ref{eq:alphbetsf}) for $\alpha$ and $\beta$ puts us into the same form as the $D$-dimensional action (\ref{stringact1}) at tree level, we immediately inherit all of the symmetry properties attributed to (\ref{stringact1}), now in $p+1$ dimensions. More precisely, the $O(D-1,D-1)$ symmetry of the parent action (\ref{stringact1}) at tree level is made manifest in the daughter action (\ref{redstringact}) for the choice (\ref{eq:alphbetsf}), exhibiting $O(p,p)$ symmetry. It is this symmetry (plus the usual time reversal invariance) which leads to interesting cosmological solutions, particularly the `pre-big bang scenario' \cite{Gasperini:1992em}. 

More explicitly, consider a Bianchi-I type metric
\beq ds^{2}=-dt^{2}+\sum_{i=1}^{p}\mathcal{R}_{i}^{2}(t)dx_{i}^{2}\;,\quad \phi=\phi(t)\;.\eeq
For this type of spacetime, the cosmological equations (arising from setting $\alpha'=0$, $A=-1$, $b_{1}=1$ and $V=0$ in (\ref{graveom}) and (\ref{phieom}))  are easy to work out and respect time reversal invariance and have a `scale factor duality' symmetry (understood to be T-duality symmetry, a subgroup of the $O(p,p)$ symmetry of the action) \cite{Gasperini:1991ak,Gasperini:1992em}:
\beq \mathcal{R}_{i}\to\frac{1}{\mathcal{R}_{i}}\;,\quad \phi\to\phi-2\sum_{i=1}^{p}\log\mathcal{R}_{i}\;.\eeq
Consequently, expanding solutions ($H_{i}\equiv\frac{\dot{\mathcal{R}}_{i}}{\mathcal{R}_{i}}>0$) are dual to contracting solutions ($H_{i}<0$). Moreover, solutions with decreasing curvature are dual to solutions with increasing curvature.\footnote{This condition depends on the signs of $\ddot{\mathcal{R}}_{i}$ and $\dot{H}_{i}$, and whether one is in an expanding or contracting phase. For example, when the decelerated expanding phase ($H_{i}>0$, $\ddot{\mathcal{R}}_{i}<0$ has decreasing curvature $\dot{H}_{i}<0$, it is dual to a decelerated contracting solution ($\dot{H}_{i}<0$, $\ddot{\mathcal{R}}_{i}>0$) with decreasing curvature $\dot{H}_{i}>0$, or an accelerated expanding solution with increasing curvature $(\dot{H}_{i}>0$, $\ddot{\mathcal{R}}_{i}>0$, $\dot{H}_{i}>0$).} This leads to the pre-big bang scenario \cite{Gasperini:1992em}: the universe reached its present state after starting (pre big bang era) in a mostly flat empty universe (the `string perturbative vacuum', where the string coupling $g_{s}^{2}=e^{\phi}\ll1$) then begins to contract with increasing curvature until reaching a state of of maximum curvature (at around the time of the big bang) and then, post big bang, expands with decreasing curvature, eventually matching standard cosmological evolution.  The above discussion is also valid when we include perfect fluid sources and a dilaton potential, however for simplicity we neglect such contributions.\footnote{When we include non-local but generally covariant potentials, the reduced $p+1$-dimensional action at tree level admits bouncing cosmologies. Explicitly, for $V(\phi)=V_{0}\exp\left(\phi-p\log \mathcal{R}\right)$ and a Bianchi-I type spacetime is assumed to be isotropic, we have an exact self-dual bouncing cosmology solution \cite{Gasperini:1996np},
$$\mathcal{R}(t)=\mathcal{R}_{0}\left[\frac{t}{t_{0}}+\left(1+\frac{t^{2}}{t_{0}^{2}}\right)^{1/2}\right]^{1/\sqrt{p}}\;,\quad \phi=-\frac{1}{2}\log\left[\sqrt{V_{0}}t_{0}\left(1+\frac{t^{2}}{t_{0}^{2}}\right)\right]+p\log\mathcal{R}(t)\;.$$}


We reiterate the self-duality symmetry of the tree-level action in $D$-dimensions is recovered in the $p+1$-dimensional theory upon choosing specific values of $\alpha$ and $\beta$. For general $\alpha,\beta$, it appears KK reduction breaks this symmetry. It is always possible, however, to perform a conformal transformation on the metric, $g_{\mu\nu}=e^{2\psi}\bar{g}_{\mu\nu}$, and move to string frame where the dilaton factor in front of the tree-level action is $e^{-\phi}$. A standard calculation shows for $\psi=-\frac{1}{(p-1)}[(p-1)\alpha+n\beta]\phi$, one finds the conformally transformed tree-level Lagrangian density is:
\beq \sqrt{-\bar{g}}e^{-\phi}[\bar{R}+b_{1}(\bar{\nabla}\phi)^{2}]\;,\quad b_{1}\equiv\frac{2p}{(p-1)}[(p-1)\alpha+n\beta]-\frac{p}{(p-1)}[(p-1)\alpha+n\beta]^{2}+a_{1}\;,\eeq
where $a_{1}$ is given in (\ref{ais}). If we then set $b_{1}=1$, we precisely arrive to the specific values of $(\alpha,\beta)$ given in (\ref{eq:alphbetsf}). In other words, the $O(D-1,D-1)$ symmetry of the tree-level action reduces to a $O(p,p)$ symmetry of the $p+1$-dimensional tree-level action. 


\subsection*{First $\alpha'$ correction}

As in the $D$-dimensional context, the reduced $p+1$-dimensional theory, at tree level, provides a viewpoint of the universe different from the standard cosmological model due to the symmetry of self-duality. Rather than evolving from a very hot, highly curved and dense initial state, as in the standard model, self-duality promotes the possibility of an initially cold and flat vacuum configuration (the pre-big bang era). Toward the end of the pre-big bang phase, the universe undergoes an accelerating growth of curvature for which there is no mechanism to exit (at tree level) and enter the standard model evolution we observe. This is in fact not an issue, as during this phase it is expected quantum effects -- including both higher curvature ($\alpha'$), and higher genus ($g^{n}_{s}$) -- will become important. This is because the growth of the curvature is accompanied by the growth of the dilaton, such that the string curvature scale is reached, $\lambda_{s}\mathcal{R}\sim1$, and, since $g_{s}\sim e^{\phi/2}$, we enter a regime of strong coupling, $g_{s}\sim1$. Thus, near the end of the pre-big bang era, the validity of the tree-level approximation breaks down and higher $\alpha'$ and genus corrections are necessary to describe a smooth transition between pre- and post-big bang epochs.

The higher derivative first $\alpha'$ correction to the $D$-dimensional effective string action (\ref{stringact1}) is important for stabilizing the maximum curvature the spacetime reaches at the big bang (when we have reached string scale). Precisely, when one works in the high curvature- weak coupling regime (neglecting quantum loop corrections) the string effects characterized solely by the $\alpha'$ correction dampen the accelerating growth such that the curvature tends to a constant value near the string scale, while the scale factor remains accelerated and the dilaton may continue to grow \cite{Gasperini:1996fu}. In particular, the specific action (\ref{stringact1}) leads to solutions of constant curvature with a linearly evolving dilaton, avoiding a big bang singularity \cite{Gasperini:1991ak}.  Such a solution is referred to as a `fixed point'.

 Naturally then, one may wonder whether the lower $(p+1)$-dimensional theory with our string frame choice (\ref{eq:alphbetsf}) exhibit a similar stabilization mechanism when we include the $\alpha'$ correction. Immediately we see the string frame choice (\ref{eq:alphbetsf}) does not lead to the same $\alpha'$ correction as given in (\ref{stringact1}). The resulting action is
\beq
\begin{split}
 I_{\text{SF}}&=-\frac{1}{2\lambda_{s}^{p-1}}\int d^{p+1}x\sqrt{-g}\biggr[e^{-\phi}[R+(\nabla\phi)^{2}]\\
&-\frac{\alpha'}{4}e^{-(2\alpha+1)\phi}\left(\mathcal{L}_{\text{GB}}+a_{2}G^{\mu\nu}(\nabla_{\mu}\phi)(\nabla_{\nu}\phi)+a_{3}(\nabla\phi)^{2}\Box\phi+a_{4}(\nabla\phi)^{4}\right)\biggr]\;,
\end{split}
\label{eq:sfact}\eeq
 whose $\alpha'$ correction is markedly different from the one appearing in (\ref{stringact1}). In particular, the problematic piece is the conformal factor multiplying the $\alpha'$ correction, $e^{-(2\alpha+1)\phi}$, such that there is not a field redefinition which can be made to bring the $\alpha'$ correction to the same form as in the parent $D$-dimensional action. This may suggest the string frame action (\ref{eq:sfact}) is unable to stabilize the curvature growth. Below we will nonetheless uncover solutions exhibiting a linearly growing dilaton with curvature tending toward a constant value, though, whether this solution provides a fixed point as described in \cite{Gasperini:1991ak} remains obscure.


\section{FLRW cosmology at $O(\a')$} \label{sec:FRLWcosmo}

Let us now explore solutions to the full $\alpha'$-corrected theory (\ref{redstringact}). We look for flat FLRW solutions when the dilaton takes a specific form. Further, as described below, we uncover and analyze solutions in various frames of interest.

To simplify notation we define the coefficient in the exponential multiplying the Ricci scalar as
\begin{equation}
f_E = (p-1)\alpha+n\beta-1\;,
\end{equation}
such that our overall $p+1$-dimensional action is written as
\begin{equation}
\begin{split}
 I&=-\frac{1}{2\lambda^{p-1}_{s}}\int d^{p+1}x\sqrt{-g}\biggr\{e^{f_E \phi}\left[R+a_{1}(\nabla\phi)^{2}\right]\\
&-\frac{\alpha'}{4}e^{(f_E-2\alpha)\phi}\left[\mathcal{L}_{GB}+a_{2}G^{\mu\nu}(\nabla_{\mu}\phi)(\nabla_{\nu}\phi)+a_{3}(\nabla\phi)^{2}\Box\phi+a_{4}(\nabla\phi)^{4}\right]\biggr\}\;.
\end{split} 
\label{eq:usefulformofact}\end{equation}
As previously mentioned, due to our freedom in choosing the values for $\a$ and $\beta$ we can define various frames in which to study the resulting solutions. There are four different frames of interest: (i) Einstein, (ii) string, (iii) dual, and (iv) Gauss-Bonnet. Each are distinguished by the form of the exponential dilaton coupling and the normalization on the standard kinetic term, whose coefficient $a_1$ is given in (\ref{ais}). More precisely, the four frames are defined as

\begin{itemize}

\item Einstein: $f_E=0$, $a_1 = -\frac{1}{2},$ yielding
\begin{equation}\label{abplus}
(\alpha_{+},\beta_{+}) = \Bigg(\frac{2p-2+\sqrt{2n(p-1)(p+n-3)}}{2(p-1)(p+n-2)}, \ \frac{2n-\sqrt{2(p-1)(p+n-3)}}{2n(p+n-1)}\Bigg),
\end{equation}
\begin{equation}\label{abminus}
(\alpha_-,\beta_-) = \Bigg( \frac{2p-2-\sqrt{2n(p-1)(p+n-3)}}{2(p-1)(p+n-1)}, \ \frac{2n+\sqrt{2n(p-1)(p+n-3)}}{2n(p+n-1)}\Bigg),
\end{equation}

\item String: $f_E=-1$, $a_1 = +1,$ yielding
\beq\label{stringframeab}
 \alpha=-\frac{2n}{(p-1)(p+n-1)}\;,\quad \beta=\frac{2}{p+n-1}\;,
\eeq

\item Dual: $\alpha = 0,$ freedom in $\beta$ 
\begin{itemize}
\item Gauss-Bonnet: $\a=0,$ $\beta = 1/n.$
\end{itemize}

\end{itemize}
Thus, the Einstein frame removes the exponential term at tree-level and canonically normalizes the dilaton kinetic energy, conditions which admit two different choices for $\alpha$ and $\beta$ due to $a_1$ being quadratic in $\alpha$; the string frame demands the reduced tree-level action matches the $D$-dimensional action, and the dual frame amounts to a single, non-trivial overall exponential factor. The Gauss-Bonnet frame results in no overall exponential factor in front of the Gauss-Bonnet term and is a special case of the dual frame. Note that if the reduced theory is in $3+1$ dimensions, the Gauss-Bonnet term is purely topological in the Gauss-Bonnet frame.

\subsection{Field equations and dilaton ansatz}

Consider now a spatially flat, homogenous and isotropic $p+1$-dimensional FLRW spacetime with line element 
\begin{equation}
ds_{p+1}^2 = -dt^2+\mR(t)^2\delta_{ij}dx^i dx^j,
\end{equation}
where $\mR(t)$ denotes the cosmological scale factor and the $i,j,...$ indices run over the $p$-spatial dimensions. The evolution of the spacetime is governed by the $(tt)$-component of the metric variation, which we refer to as the Friedmann equation, and the equation of motion for the dilaton. The full covariant field equations are rather cumbersome and are given in (\ref{graveom}) and (\ref{phieom}), respectively. Here we further assume an isotropic dilaton, $\phi=\phi(t)$, which significantly simplifies the background equations. The Friedmann equation  becomes
\begin{equation}\label{Fried1}
\begin{split}
0&=e^{f_E \phi}\mR^{p-2}\Big[p(p-1)\mRd^2 +2pf_E \mR\mRd\pd+a_1\mR^2\pd^2 \Big] \\
&-\frac{\alpha'}{4}e^{(f_E-2\alpha)\phi}\mR^{p-4}\Bigg[p(p-1)(p-2)(p-3)\mRd^4+4p(p-1)(p-2)(f_E-2\alpha)\mR\mRd^3\pd \\
&-\frac{3}{2}p(p-1)a_2\mR^2\mRd^2\pd^2 
-2pa_3\mR^3\mRd\pd^3+((f_E-2\alpha)a_3-3a_4)\mR^4\pd^4\Bigg],
\end{split}
\end{equation}
while the equation of motion for the dilaton is
\begin{equation}\label{phi1}
\begin{split} 
0&=e^{f_E\phi}\mR^{p-2}\Big[p(p-1)f_E\mRd^2+2p\mR(a_1\mRd\pd+f_E\mRdd)+a_1\mR^2(f_E\pd^2+2\pdd)\Big] \\
&-\frac{\alpha'}{4}e^{(f_E-2\alpha)\phi}\mR^{p-4}\Bigg[p(p-1)(p-2)(p-3)(f_E-2\alpha)\mRd^4-p(p-1)(p-2)a_2\mR\mRd^3\pd \\
&-\mR^3\pd^2\Big(2pa_3\mRdd-\mR((f_E-2\alpha)a_3-3a_4)((f_E-2\alpha)\pd^2+4\pdd)\Big) \\
&-\frac{1}{2}p(p-1)\mR\mRd^2\Big(8(2-p)(f_E-2\alpha)\mRdd+\mR(((f_E-2\alpha)a_2+4a_3)\pd^2+2a_2\pdd)\Big) \\
&-2p\mR^2\mRd\pd\Big(a_2(p-1)\mRdd+2\mR(a_4\pd^2+a_3\pdd)\Big)\Bigg].
\end{split}
\end{equation}
Though (\ref{Fried1}) and (\ref{phi1}) appear formidable, we will nonetheless be able to find exact solutions.





Before we move on to specific frames, \emph{i.e.}, specific choices of ansatz coefficients $\alpha$ and $\beta$, note that one class of exact solutions is obtained by assuming the dilaton's profile to be of the form
\begin{equation}\label{phiansatz}
\phi(t) = q\log\mR(t),
\end{equation}
where $q$ is a real constant to be determined. This is a standard ansatz which is used to look for FLRW spacetimes with scale factor that generically goes as $\mR(t)\sim t^{\eta}$, where $\eta$ is typically not an integer value. Equation (\ref{phiansatz}) will also encompass a second typical class of solutions: constant curvature spacetime $\mR(t)\sim e^{H_0 t}$ and corresponding linear dilaton.

 Feeding (\ref{phiansatz}) through (\ref{Fried1}) yields
\begin{equation}\label{Free}
0=e^{f_E\phi}\mR^{p-2}\mRd^2\Bigg[P_0 - \alpha' P_1\Bigg(\frac{\mRd}{\mR^{1+\a q}}\Bigg)^2\Bigg],
\end{equation}
where the polynomials are given by
\begin{equation}\label{newP0}
P_0(q) = p(p-1)+2pqf_E+a_1 q^2,
\end{equation}
\begin{equation}\label{newP1}
\begin{split}
P_1(q) &= \frac{1}{4}\Big[p(p-1)(p-2)(p-3)+4p(p-1)(p-2)(f_E-2\a)q-\frac{3}{2}p(p-1)a_2 q^2 \\
& \ \ \ \ \ \ \ -2pa_3 q^3+[(f_E-2\a)a_3-3a_4]q^4\Big].
\end{split}
\end{equation}
Using (\ref{phiansatz}), we can next simplify (\ref{phi1}) to
\begin{equation}\label{phi3}
\begin{split}
0=e^{f_E\phi}\mR^p\Bigg[A\Big(\frac{\mRd}{\mR}\Big)^2+B\frac{\mRdd}{\mR}-\a'\mR^{-2\a q}\Bigg(C\Big(\frac{\mRd}{\mR}\Big)^4-D\Big(\frac{\mRd}{\mR}\Big)^2\Big(\frac{\mRdd}{\mR}\Big)\Bigg)\Bigg],
\end{split}
\end{equation}
where 
\begin{equation}
\begin{split}
A &= p(p-1)f_E+2a_1pq+a_1f_Eq^2-2a_1 q, \\
B &= 2pf_E+2a_1q, \\
C&= \frac{1}{4}\Big[p(p-1)(p-2)\Big((p-3)(f_E-2\a)-a_2 q\Big)+q^2\Big((f_E-2\a)a_3-3a_4\Big)\Big(q^2(f_E-2\a)-4q\Big) \\
& \ \ \ \ \ -\frac{1}{2}p(p-1)\Big(q^2\big((f_E-2\a)a_2+4a_3\big)-2a_2 q\Big) -2pq\big(2a_4 q^2-2a_3 q\big)\Big], \\
D &=\frac{1}{4}\Big[ q^2\Big(2pa_3-4q\big((f_E-2\a)a_3-3a_4\big)\Big)+\frac{1}{2}p(p-1)\big(8(2-p)(f_E-2\a)+2a_2 q\big) \\
& \ \ \ \ \ +2pq\big(a_2(p-1)+2a_3 q\big)\Big].
\end{split}
\end{equation}
The approach is now to use (\ref{Free}) to determine the scale factor, and then use (\ref{phi3}) to fix the value of $q$. We will do so for specific frames, starting with the Einstein and string frame.

\subsection{Einstein and string frame solutions}

Both the Einstein frame and string frame share the conditions that $\a\neq 0$ and $\beta\neq 0.$ Because of this, the general structure of the solutions associated to these frames is the same. We begin by noting that a trivial solution to (\ref{Free}) is when $\mathcal{R}(t)$ is a constant, \emph{i.e.}, the Minkowski space solution, which we will ignore in our analysis.  In the infinite string tension limit $\alpha'\to0$, (\ref{Free}) is satisfied when $P_{0}(q)=0$,  which, in the Einstein frame where $f_E=0,$ sets $q=\pm\sqrt{2p(p-1)}$ so that $\phi(t)=\pm\sqrt{2p(p-1)}\log\mR(t)$. The scale factor is then determined by (\ref{phi3}), with the solution $\mR(t)=c_0(t-t_0)^{1/p}$. Thus we recover the `stiff fluid' cosmology \cite{Banks:2004cw,Banks:2004vg,Banks:2003ta} (see also \cite{Battefeld:2004jh}) where a flat FLRW universe is dominated by a fluid with the stiff equation $\rho=P$ (for example, a dense black hole fluid).\footnote{The stiff fluid model is an example of a holographic cosmology.  In particular, near the big bang singularity small causal diamonds have few degrees of freedom and a non-singular quantum description, a consequence of Bousso's general formulation of the holographic principle and the covariant entropy bound \cite{Bousso:1999cb}.} As we will see, however, the general solution for $\alpha'\neq0$ is unable to smoothly connect to the stiff fluid universe.

Keeping $\alpha'\neq0$ and integrating (\ref{Free}) from an arbitrary initial time $t_0$, we find
\begin{equation}\label{generalscalefactor}
\mathcal{R}(t)=c_0(t-\tilde{t}_0)^{\eta},
\end{equation}
with 
\begin{equation}\label{c0t0}
\eta\equiv -\frac{1}{\alpha q}, \ \ \ \ \ \ \ 
c_0 = \Bigg(\frac{\epsilon\alpha q}{\sqrt{\alpha'}}\sqrt{\frac{P_0(q)}{P_1(q)}}\Bigg)^{\eta}, \ \ \ \ \ \ \ \tilde{t}_0 = t_0-\frac{\sqrt{\alpha'}}{\epsilon\alpha q}\sqrt{\frac{P_1(q)}{P_0(q)}}\mR_0^{-\alpha q}\;.
\end{equation}
Here $\mR_0\equiv \mR(t_0)$ and $\epsilon=\pm1$ comes from taking the square root of the term in the parenthesis in (\ref{Free}). In principle, either choice of $\epsilon$ represents a valid solution. In Einstein gravity, $\epsilon=+1$ corresponds to cosmological expansion while $\epsilon=-1$ corresponds to contraction, however, for non-minimally coupled theories such as the Horndeski theory under consideration, this may not be the case. As we will discuss later, for the solutions under consideration the sign of $\epsilon$ will drop out of our expression for the Hubble parameter.  At time $t=\tilde{t}_{0}$ the expansion factor vanishes for nonnegative $\eta$, signaling the time of the big bang, where pre-big bang and post-big bang epochs occur when $t<\tilde{t}_{0}$ and $t>\tilde{t}_{0}$, respectively. Lastly, we observe the singular limit $\alpha'\to0$ (when $\eta>0$), such that the $\alpha'\neq0$ solution does not smoothly connect to the stiff fluid universe $\alpha'=0$.

The next step is to utilize the dilaton equation of motion to determine the allowed values of $q$. It is useful to note that (\ref{generalscalefactor}) implies
\begin{equation}
\frac{\mRd}{\mR}=\frac{\eta}{t-\tilde{t}_0}, \ \ \ \ \ \ \ \frac{\mRdd}{\mR}=\frac{\eta(\eta-1)}{(t-\tilde{t}_0)^2},
\end{equation}
along with 
\begin{equation}\label{Rtomaq}
\mR^{-2\a q}=\frac{\a^2 q^2 P_0}{\a' P_1}(t-\tilde{t}_0)^2.
\end{equation}
The relation (\ref{Rtomaq}) allows us to factor all of the time-dependence out of the equation of motion for the dilaton (\ref{phi3}), reducing it to a polynomial in $q$:
\begin{equation}
0=\frac{\mR^p\eta^2}{P_1(t-\tilde{t}_0)^2}\mathcal{P}_{p,n}(q),
\end{equation}
where, after using $\eta=-1/\a q,$ 
\begin{equation}\label{EinsteinPee}
\mathcal{P}_{p,n}(q)\equiv P_1(q)\Big(A+B(1+\a q)\Big)-P_0(q)\Big(C-D(1+\a q)\Big). 
\end{equation}
This polynomial is in general fifth order in $q$ whose coefficients nontrivially depend on the choice of the internal and bulk spatial dimensions.

The roots of (\ref{EinsteinPee}) provide the values of $q$ that satisfy the field equations, however, we must be cautious in ensuring the dilaton and metric are not complex. Suppose for a given $p,n$, $q^*$ is such that $\mathcal{P}_{p,n}(q^*)=0.$ Then for the dilaton as in (\ref{phiansatz}) to be real, $q^*$ clearly must be real. A second reality condition comes from the integration of the Friedmann equation; the constant $c_0$ must have at least one real principal root for the scale factor $\mR(t)$ to be real. This is guaranteed when the ratio $\sqrt{\frac{P_0(q^*)}{P_1(q^*)}}\in\mathbb{R}$, as should be apparent from (\ref{c0t0}). Thus for any root $q^*$, we must verify 
\begin{equation}\label{realcondish}
\text{I:} \ \ \ \ \ q^*\in\mathbb{R}, \  \ \ \ \ \ \ \ \ \text{II:} \ \ \ \ \ \ \frac{P_0(q^*)}{P_1(q^*)}>0.
\end{equation}

Determining the valid roots (as well as the growth factor $\eta$) requires specifying the internal and bulk spatial dimensions $n$ and $p$. The growth factor will take on different values in each dimension, and in some dimensions, the ansatz (\ref{phiansatz}) will not admit solutions corresponding to a real dilaton and scale factor in the two frames under consideration.  Additionally, due to $\a$ and $\beta$ taking on differing values between the Einstein and string frames, the roots and growth factor $\eta$ will not be the same. For our analysis, we assume that the total dimension $D = p+1+n$ is $D=10,$ the critical dimension for the heterotic string. In Appendix \ref{AppendixForRates}, we illustrate the process of obtaining the value for the growth factor for both the Einstein and string frames for the case of three bulk spatial dimensions ($p=3, n=6$). The results are summarized for the expansion rates for other relevant values of $p$ in Tables \ref{plussolstable}, \ref{minussolstable2}, and \ref{stringtable} below (we do not consider trivial roots $q^*=0$).

Interesting points to note are that the Einstein frame solutions associated to $(\a_+,\beta_+)$ all come along with two valid roots and therefore two solutions for the growth factor $\eta$. In 3+1 dimensions, we see that both values for $\eta$ are such that $\eta >2,$ which is not the case for $p\geq 4.$ Loosely speaking, $\ddot{\mR}>0$ corresponds to inflationary-type expansion, which occurs whenever $\eta\geq 2.$  The Einstein frame solutions associated to the $(\a_-,\beta_-)$ are in general quite odd in that their growth rates are significantly larger than the power law solutions produced by standard perfect fluids.\footnote{For example, it is well known that the radiation dominated universe generically grows as $\mR(t)\sim t^{1/2}$ while the matter dominated universe grows as $\mR(t)\sim t^{2/3}$.}  Moreover, since $\a_-=0$ in 7+1 dimensions, the growth rate $\eta=-1/\a q^*$ is ill defined, although no valid roots exist in that case. The solutions in string frame have growth rates that are also significantly larger than `normal'. In $p=6,7,8$, $\eta<0$ indicates singular behavior as $t\rightarrow\infty$, suggesting the dimension of spacetime is either four or five dimensional (as universes in the other dimensions would quickly decay).  It is our expectation that these undesirable features are related to the coupling regime in which a given solution resides. 


\begin{table}[t]
\begin{center}
\begin{tabular}{c| c | c | c | c }
$p,n$ & $(\alpha_+,\beta_+)$  & Real roots & Valid roots & $\eta$ \\ \hline 
2, 7 & $\Big(\frac{1+\sqrt{21}}{8},\frac{7-\sqrt{21}}{8}\Big)$ & 3 & 0 & $\sim$ \\
3, 6 & (1/2,0) & 3 & 2 & 2.85, 3.42 \\
4, 5 & $\Big(\frac{1+\sqrt{5}}{8},\frac{5-3\sqrt{5}}{40}\Big)$ &  3& 2 & 1.87, 3.93 \\
5, 4 & $\Big(\frac{1+\sqrt{3}}{8}, \frac{1-\sqrt{3}}{8}\Big)$ & 3& 2 & 1.56, 4.38 \\
6, 3 & $\Big(\frac{5+3\sqrt{5}}{40}, \frac{1-\sqrt{5}}{8}\Big)$ &  3 & 2 & 1.41, 4.97  \\
7, 2 & $(1/4,-1/4)$  &3 & 2 & 1.36, 5.89 \\
8, 1 & $\Big(\frac{7 + \sqrt{21}}{56}, \frac{1-\sqrt{21}}{8}\Big)$  & 3 & 2 & 1.43, 7.62
\end{tabular}
\end{center}
\caption{Einstein frame solutions associated to $(\a_+,\beta_+)$ as in (\ref{abplus}) in various dimensions. The last column on the right summarizes the rate of growth for each dimension. }\label{plussolstable}
\end{table}


\begin{table}[H]
\begin{center}
\begin{tabular}{c| c | c | c | c }
$p,n$ & $(\alpha_-,\beta_-)$ & Real roots & Valid roots & $\eta$ \\ \hline 
2, 7 & $\Big(\frac{1-\sqrt{21}}{8},\frac{7+\sqrt{21}}{8}\Big)$  & 3 & 0& $\sim$ \\
3, 6 & (-1/4,1/4) & 3 & 1 & 35.7 \\
4, 5 & $\Big(\frac{1-\sqrt{5}}{8},\frac{5+3\sqrt{5}}{40}\Big)$ &  3 & 1 & 49.9 \\
5, 4 & $\Big(\frac{1-\sqrt{3}}{8}, \frac{1+\sqrt{3}}{8}\Big)$ &  3 & 1 & 96.7 \\
6, 3 & $\Big(\frac{5-3\sqrt{5}}{40}, \frac{1+\sqrt{5}}{8}\Big)$ &  3 & 1 & 333.5 \\
7, 2 & $(0,1/2)$ & 3 & 0 & $\sim$ \\
8, 1 & $\Big(\frac{7 - \sqrt{21}}{56}, \frac{1+\sqrt{21}}{8}\Big)$ & 3 & 1 &  213.4
\end{tabular}
\end{center}
\caption{Einstein frame solutions associated to $(\a_-,\beta_-)$ as in (\ref{abminus}) in various dimensions. Of note, in $7+1$ dimensions the allowed solutions would lead to ill-defined scale factor growth, since $\a_-=0$, although no roots to $\mathcal{P}^{-}_{7,2}$ are valid. }
\label{minussolstable2}\end{table}

\begin{table}[H]
\begin{center}
\begin{tabular}{c| c | c | c | c }
$p,n$ & $(\alpha,\beta)$ & Real roots & Valid roots & $\eta$ \\ \hline 
2, 7 & (-7/4, 1/4)  & 3 & 0& $\sim$ \\
3, 6 & (-1/4, 1/4) & 3 & 1 & 12.6 \\
4, 5 & (-5/12, 1/4) &  3 & 1 & 31.7 \\
5, 4 & (-1/4, 1/4) &  3 & 0 & $\sim$ \\
6, 3 & (-3/20, 1/4) &  3 & 1 & -62.4 \\
7, 2 & (-1/12, 1/4) & 3 & 2 & -1.7, -50.6 \\
8, 1 & (-1/28, 1/4) & 3 & 2 &  -3.9, -72.3
\end{tabular}
\end{center}
\caption{String frame solutions associated to $(\a,\beta)=\big(\frac{-2n}{(p-1)(p+n-1)},\frac{2}{p+n-1}\big)$ in various dimensions. }
\label{stringtable}\end{table}


\subsection{Dual frame solutions}

Both the Dual and Gauss-Bonnet frames share $\a=0,$ which, by inspection to the Friedmann equation (\ref{Free}), results in 
\begin{equation}\label{Fried3}
0=e^{f_E\phi}\mR^{p-2}\mRd^2\Bigg[\tilde{P}_0(q)-\a'\tilde{P}_1(q)\Big(\frac{\mRd}{\mR}\Big)^2\Bigg]\;,
\end{equation}
where the structure of $\tilde{P}_0,\tilde{P}_1$  (\ref{newP0}) and (\ref{newP1}) is now slightly different from the Einstein and string frame cases. The nontrivial solution to (\ref{Fried3}) is (anti) de Sitter space,
\begin{equation}\label{ADS1}
\mR(t) = \mR_0 e^{H_0 t}, \ \ \ \ \ \ \ \ H_0 = \frac{\epsilon}{\ell_s}\sqrt{\frac{\tilde{P}_0(q^*)}{\tilde{P}_1(q^*)}},
\end{equation}
where again $\epsilon=\pm 1$ and we have explicitly written the string scale $\ell_s=\sqrt{\a'}$. Similar to the cases previously examined, $q^*$ will be determined by the field equation of the dilaton. In fact, the equation of motion for the dilaton reduces to a polynomial in $q$, the likes of (\ref{EinsteinPee}) and (\ref{stringpoly}), but with $\a\rightarrow 0$:
\begin{equation}\label{tildeBigP}
\tilde{\mathcal{P}}_{p,n}(q)=\tilde{P}_1(q)(\tilde{A}+\tilde{B})-\tilde{P}_0(q)(\tilde{C}-\tilde{D}),
\end{equation}
where 
\begin{equation}\label{tildeP0}
\tilde{P}_0(q) = p(p-1)+2pq(n\beta-1)+\tilde{a}_1 q^2,
\end{equation}
\begin{equation}\label{tildeP1}
\begin{split}
\tilde{P}_1(q) &= \frac{1}{4}\Big[p(p-1)(p-2)(p-3)+4p(p-1)(p-2)(n\beta-1)q-\frac{3}{2}p(p-1)\tilde{a}_2 q^2 \\
& \ \ \ \ \ \ \ -2p\tilde{a}_3 q^3+[(n\beta-1)\tilde{a}_3-3\tilde{a}_4]q^4\Big],
\end{split}
\end{equation}
\begin{equation}
\begin{split}
\tilde{A} &= p(p-1)(n\beta-1)+2\ta_1pq+\ta_1(n\beta-1)q^2-2\ta_1 q, \\
\tilde{B}&= 2p(n\beta-1)+2\ta_1q, \\
\tilde{C}&= \frac{1}{4}\Big[p(p-1)(p-2)\Big((p-3)(n\beta-1)-\ta_2 q\Big)+q^2\Big((n\beta-1)\ta_3-3\ta_4\Big)\Big(q^2(n\beta-1)-4q\Big) \\
& \ \ \ \ \ -\frac{1}{2}p(p-1)\Big(q^2\big((n\beta-1)\ta_2+4\ta_3\big)-2\ta_2 q\Big) -2pq\big(2\ta_4 q^2-2\ta_3 q\big)\Big], \\
\tilde{D} &=\frac{1}{4}\Big[ q^2\Big(2p\ta_3-4q\big((n\beta-1)\ta_3-3\ta_4\big)\Big)+\frac{1}{2}p(p-1)\big(8(2-p)(n\beta-1)+2\ta_2 q\big) \\
& \ \ \ \ \ +2pq\big(\ta_2(p-1)+2\ta_3 q\big)\Big].
\end{split}
\end{equation}
In the above, the $\ta_i$ are given by (\ref{ais}) after taking the limit $\a\rightarrow 0.$ Importantly, notice there is a similar reality condition to (\ref{realcondish}) on the Hubble parameter as in (\ref{ADS1}). For $q^*$ solving the equations of motion for the dilaton (\ref{tildeBigP}), we must have that 
\begin{equation}\label{realC2}
\frac{\tilde{P}_0(q^*)}{\tilde{P}_1(q^*)} > 0
\end{equation}
for the Hubble parameter to be real. This will again rule out certain real roots to (\ref{tildeBigP}), as in the other frames discussed previously.


In order to completely solve the field equations, we now must pick the number of bulk spatial dimensions as well as value for $\beta$. Let us begin by commenting on the Gauss-Bonnet frame, where $\beta=1/n.$ Interestingly, for bulk spatial dimensions $p=2,...,8,$ no real roots to (\ref{tildeBigP}) satisfy (\ref{realC2}). For example in $p=4$, where the Gauss-Bonnet combination nontrivially contributes to the field equations, (\ref{tildeBigP}) becomes
\begin{equation}
\tilde{\mathcal{P}}_{4,5}(q) = -\frac{2q}{625}\big(53 q^4-1920 q^3-840 q^2-115200q-213000\big).
\end{equation}
There are two real roots in addition to the trivial $q^*=0,$ which are $q^*_a\approx -1.77, \  q^*_b\approx 38.2.$ Checking (\ref{realC2}) shows $\frac{\tilde{P}_0(q_a^*)}{\tilde{P}_1(q_a^*)}<\frac{\tilde{P}_0(q_b^*)}{\tilde{P}_1(q_b^*)} <0,$ signaling that there are no real solutions in the Gauss-Bonnet frame when $p=4.$ It is straightforward to numerically solve the roots of (\ref{tildeBigP}) and check the condition (\ref{realC2}) in other dimensions, and one finds that (\ref{realC2}) is never satisfied. We will expand on this further shortly.

In the dual frame, however, we indeed find real solutions to the field equations in difference to the Gauss-Bonnet frame as a result of our freedom to choose $\beta.$ This freedom begs the question of which, if any, values for $\beta$ provide solutions consistent with phenomenological considerations. One way to gain access to that answer is by examining the stability of the metric perturbations for general $\beta,$ and demand that the perturbations be ghost free as well as $c_{gw}\equiv 1$ and $0<c_s\leq 1,$ where $c_{gw}$ and $c_s$ are the propagation speeds for the tensor and scalar perturbations, respectively. Those demands turn out to be too stringent for the dual frame solutions. In the following section, focusing on the tensor perturbations, we will discuss how only certain values for $\beta$ permit ghost free solutions and, interestingly, such solutions also generically are superluminal in that $c_{gw}^2>1.$


\section{Stability in 3 + 1 dimensions} \label{sec:stability}

Above we found various FLRW solutions in a variety of frames, including exact solutions exhibiting cosmic acceleration. Naturally, one may be interested in whether cosmological solutions are stable under metric perturbations.  The question of stability of a Horndeski theory of the type (\ref{Horndaction}) on a FLRW background was first analyzed in \cite{DeFelice:2011bh}, in which general conditions on the functions $K,G_{i}$ were given to maintain a ghost free theory and rendered stable under scalar and tensor perturbations. Specifically, the squared propagation speed of tensor perturbations $c_{gw}^{2}$ in a FLRW background is given by (reviewed in Appendix \ref{app:linperts})  
\beq c^{2}_{gw}\equiv\frac{G_{4}-X(\ddot{\phi}G_{5,X}+G_{5,\phi})}{G_{4}-2XG_{4,X}-X(H\dot{\phi}G_{5,X}-G_{5,\phi})}\;.\label{eq:gwspeedgen}\eeq
The expression for the scalar sound speed is significantly more complicated and is left for the Appendix. In the case of general relativity plus a scalar field, $G_{2}=X-V(\phi)$, $G_{4}=1$, $G_{3}=G_{5}=0$, one has $c_{gw}^{2}=1$, \emph{i.e.}, tensor perturbations propagate at the speed of light. More generally $c_{gw}^{2}\neq1$ such that speeds $c_{gw}^{2}<1$ or $c^{2}_{gw}>1$ are said to be subluminal or superluminal, respectively. Theories admitting modes of any kind propagating superluminally are generally considered undesirable as they seem to lead to potential paradoxes due to violations in causality, loss of analyticity of the S-matrix and UV incompleteness (see, \emph{e.g.}, \cite{Adams:2006sv,Shore:2007um} in the context of theories of k-essence, and, for more recent accounts, \emph{e.g.}, \cite{Dobre:2017pnt,deRham:2021fpu}). Thus, \emph{a priori} subluminal propagation is preferred and sometimes used as a way to impose restrictions on various effective field theories. We point out, however, the presence of external matter can influence superluminality and hence properties of stability \cite{Easson:2013bda}.

It is worthwhile then to consider whether the FLRW solutions uncovered here in the string induced Galileons remain stable and avoid superluminal propagation. Carrying out the analysis in general $p+1$ dimensions with a Gauss-Bonnet contribution is generally difficult and beyond the scope of this article. However, we can make progress if we restrict to $3+1$ dimensions, since in this case it is known a non-minimally coupled Gauss-Bonnet term of the form $\int d^{4}x\sqrt{-g}\frac{1}{8}f(\phi)\mathcal{L}_{\text{GB}}$, for some function of the dilaton $f(\phi)$, results in the following Horndeski form functions \cite{Kobayashi:2011nu,DeFelice:2011uc}:
\beq 
\begin{split}
&K^{\text{GB}}(\phi,X)=f^{(4)}(\phi)X^{2}(3-\log X)\;,\\
&G^{\text{GB}}_{3}(\phi,X)=\frac{1}{2}f^{(3)}(\phi)X(7-3\log X)\;,\\
&G^{\text{GB}}_{4}(\phi,X)=\frac{1}{2}f^{(2)}(\phi)X(2-\log X)\;,\\
&G^{\text{GB}}_{5}(\phi,X)=-\frac{1}{2}f^{(1)}(\phi)\log X\;,
\end{split}
\eeq
with $X=-\frac{1}{2}(\nabla\phi)^{2}$ and $f^{(n)}(\phi)$ denoting the $n$th derivative of $f(\phi)$ with respect to $\phi$. Adding these to the form functions (\ref{eq:formfactshorn}) of the generic string induced Horndeski action (\ref{eq:usefulformofact}), 
\beq 
\begin{split}
&K(\phi,X)=-2a_1e^{f_{E}\phi}X-\alpha'a_4e^{(f_{E}-2\alpha)\phi}X^2-2\alpha'X^{2}(f_{E}-2\alpha)^{4}e^{(f_{E}-2\alpha)\phi}(3-\log X)\;,\\
&G_{3}(\phi,X)= -\frac{\alpha'}{2}a_3e^{(f_{E}-2\alpha)\phi}X-\alpha'(f_{E}-2\alpha)^{3}e^{(f_{E}-2\alpha)\phi}X(7-3\log X)\;,\\
&G_{4}(\phi,X)=e^{f_{E}\phi}-\alpha' (f_{E}-2\alpha)^{2}e^{(f_{E}-2\alpha)\phi}X(2-\log X)\;,\\
&G_{5}(\phi,X)=\frac{\alpha'}{4}\frac{a_2}{(f_{E}-2\alpha)}e^{(f_{E}-2\alpha)\phi}+\alpha'(f_{E}-2\alpha)e^{(f_{E}-2\alpha)\phi}\log X\;.
\end{split}
\label{eq:formfuncsinduc4d}\eeq
where we used $f(\phi)=-2\alpha' e^{(f_{E}-2\alpha)\phi}$. Then,  using functions (\ref{eq:GTgen}) and (\ref{eq:FTgen}), we have  
\begin{equation}\label{GTandFT}
\begin{split}
\mathcal{G}_T &=\frac{1}{2}e^{f_E\phi}\Big[4-\a' e^{-2\a\phi}\Big(4H\dot{\phi}(f_E-2\a)-a_2X\Big)\Big], \\
\mathcal{F}_T &= \frac{1}{2}e^{f_E\phi}\Big[4-\a' e^{-2\a\phi}\Big(a_2 X+4(f_E-2\a)(\ddot{\phi}+2X(f_E-2\a))\Big)\Big],
\end{split}
\end{equation}
and
\beq \label{generalcgw} c^{2}_{gw}=\frac{4-\a' e^{-2\a\phi}\Big(a_2 X+4(f_E-2\a)(\ddot{\phi}+2X(f_E-2\a))\Big)}{4-\a' e^{-2\a\phi}\Big(4H\dot{\phi}(f_E-2\a)-a_2X\Big)} .
\eeq
For the tensor perturbations to be ghost free, $\mathcal{G}_T>0$ and $\mathcal{F}_{T}>0,$ while $c_{gw}^2>0$ ensures absence of the gradient instability. For the scalar modes, there are similar conditions on coefficients $\mathcal{F}_S, \ \mathcal{G}_S,$ and $c_s^2=\mathcal{F}_S/\mathcal{G}_S$ given in (\ref{Sigma}) - (\ref{FsGs}), where stability of the scalar perturbations implies $\mathcal{F}_S>0, \mathcal{G}_S>0$, and $c_s^2>0.$

Immediately we see at tree-level $\alpha'=0$ one has $c^{2}_{gw}=1$. For $\alpha'\neq0$, however, one can imagine using positivity of $c^{2}_{gw}$ and subluminality to constrain possible values of parameters $\alpha,\beta$. However in each frame except the dual frame, $\a$ and $\beta$ are particular values, thus we can simply check whether or not the solution is stable. 

Let us start by commenting on the Gauss-Bonnet frame, where we found that there were no real solutions in any dimensions. For the case of 3+1 dimensions ($\alpha=0,\ \beta=1/6$ and $a_2=14/3$),  (\ref{generalcgw}) becomes
\begin{equation}
c_{gw}^2=\frac{12+7\a'\dot{\phi}^2}{12-7\a'\dot{\phi}^2}.
\end{equation}
When $|\dot{\phi}|<\sqrt{\frac{12}{7\a'}},$ $c_{gw}^2>1$, while for $|\dot{\phi}|>\sqrt{\frac{12}{7\a'}},$ $c_{gw}^2<0.$ Although this result does not \emph{a priori} suggest the absence of real solutions in the Gauss-Bonnet frame, further investigation of that particular model (without additional matter fields or a dilaton potential) is unnecessary.

In the Einstein and string frames, interestingly, all of the time dependence falls out of both terms in (\ref{GTandFT}) -- and therefore also in (\ref{generalcgw}) -- resulting in $\mathcal{F}_T$, $\mathcal{G}_T$, and $c_{gw}^2$ being constant values depending solely on the root(s) to (\ref{EinsteinPee}), $q^*$, and the constants $\a$ and $\beta.$ For the case of 3+1 dimensions, we have the two parameter sets $(\a_+,\beta_+)=(1/2,0)$ and $(\a_-,\beta_-)=(-1/4,1/4)$, while in the string frame, $(\a,\beta)= (-3/4,1/4).$ There are two valid solutions admitted in the $(\a_+,\beta_+)$ model, one valid solution in the $(\a_-,\beta_-)$ model, and one valid solution in the string frame (see Tables \ref{plussolstable}, \ref{minussolstable2}, and \ref{stringtable}). For all four of these exact solutions, it is straightforward to compute (\ref{GTandFT}) and (\ref{generalcgw}), and the result is
\begin{equation}
\mathcal{F}_{T} > 0, \ \ \ \ \ \ \mathcal{G}_T < 0, \ \ \ \ \ \ \ c_{gw}^2<0.
\end{equation}
This signals that the tensor perturbations are not under control and the solutions are unstable. However, as we will illustrate in section \ref{sec:potential}, the model can be stabilized by including a dilaton potential.

To use stability conditions to constrain the value of $\beta$ in the dual frame (where $\a=0$), we took the approach of solving the field equations for a wide range in $\beta$ and checking the simplest expressions, mainly $\mathcal{F}_T, \ \mathcal{G}_T,$ and $c_{gw}^2$, associated to the particular solution. We found that various ranges of $\beta<0$ resulted in $\mathcal{F}_T>0$ and $\mathcal{G}_T>0,$ however, the propagation speed of the gravitational waves is in general superluminal. A simple example is the choice $\beta=-1,$ which, in 3+1 dimensions, results in the Hubble constant being $H_0\approx \frac{0.294}{\ell_s}$ in our units (see Table \ref{betaminustable} in Appendix \ref{appen:dualframe}). The speed of gravitational waves associated with this exact solution is $c_{gw}^2\approx 1.54.$ Moreover, taking the solution associated with $\beta=-1$ and evaluating the expressions for $\mathcal{F}_S,$ $\mathcal{G}_S$ results in $\mathcal{G}_S>0$ but $\mathcal{F}_S<0$ and thus $c_s^2<0.$ Therefore we conclude that although the tensor perturbations are ghost-free, they exhibit superluminal propagation speed and the scalar perturbations are plagued with ghosts and gradient instabilities. Consequently, none of the exact solutions we have discussed thus far are stable to all metric perturbations.

\subsection{Numerical solutions including a dilaton potential}\label{sec:potential}

We will now show that by including a potential for the dilaton $V(\phi)$, we can ensure the metric perturbations are well behaved. The inclusion of a potential results in the necessity to numerically integrate the equations of motion opposed to the exact, analytical solutions discussed previously.  After adding the potential to the $p+1$-dimensional action (\ref{eq:usefulformofact}), in terms of the isotropic dilaton and scale factor the Friedmann equation (\ref{Fried1}) becomes
\begin{equation}\label{FriedWithPot}
\begin{split}
0&=e^{f_E \phi}\mR^{p-2}\Big[p(p-1)\mRd^2 +2pf_E \mR\mRd\pd+\mR^2\big(a_1\pd^2-V(\phi)\big) \Big] \\
&-\frac{\alpha'}{4}e^{(f_E-2\alpha)\phi}\mR^{p-4}\Bigg[p(p-1)(p-2)(p-3)\mRd^4+4p(p-1)(p-2)(f_E-2\alpha)\mR\mRd^3\pd \\
&-\frac{3}{2}p(p-1)a_2\mR^2\mRd^2\pd^2 
-2pa_3\mR^3\mRd\pd^3+((f_E-2\alpha)a_3-3a_4)\mR^4\pd^4\Bigg],
\end{split}
\end{equation}
while the equation of motion for $\phi$ is modified to be 
\begin{equation}\label{phiWithPot}
\begin{split} 
0&=e^{f_E\phi}\mR^{p-2}\Big[p(p-1)f_E\mRd^2+2p\mR(a_1\mRd\pd+f_E\mRdd)+\mR^2\big(a_1 f_E\pd^2+2a_1\pdd-f_E V(\phi)-\frac{dV}{d\phi}\big)\Big] \\
&-\frac{\alpha'}{4}e^{(f_E-2\alpha)\phi}\mR^{p-4}\Bigg[p(p-1)(p-2)(p-3)(f_E-2\alpha)\mRd^4-p(p-1)(p-2)a_2\mR\mRd^3\pd \\
&-\mR^3\pd^2\Big(2pa_3\mRdd-\mR((f_E-2\alpha)a_3-3a_4)((f_E-2\alpha)\pd^2+4\pdd)\Big) \\
&-\frac{1}{2}p(p-1)\mR\mRd^2\Big(8(2-p)(f_E-2\alpha)\mRdd+\mR(((f_E-2\alpha)a_2+4a_3)\pd^2+2a_2\pdd)\Big) \\
&-2p\mR^2\mRd\pd\Big(a_2(p-1)\mRdd+2\mR(a_4\pd^2+a_3\pdd)\Big)\Bigg].
\end{split}
\end{equation}
The above two equations are what we will be numerically integrating after setting $p=3,n=6,$ and, for simplicity, we will focus on the Einstein frame and set $(\a,\beta)$ to $(\a_+,\beta_+)=(1/2,0)$.

In terms of the Horndeski functions (\ref{eq:formfuncsinduc4d}), the only modification at the level of the action is the addition of the potential term in $K(\phi,X)$. It's worth noting that the presence of the potential does not change the form of any of the tensor or scalar perturbation quantities because no derivatives of $K$ with respect to $\phi$ appear in the calculation of the general results. Thus the role of $V(\phi)$ is solely to modify the structure of the solutions.

For concreteness, we will consider a Higgs-like potential of the form
\begin{equation}\label{Vhiggs}
V(\phi) = (\phi^2-\lambda^2)^2.
\end{equation}
In the standard cosmological scenario, a scalar field in an expanding spacetime experiences Hubble friction and tends to oscillate about a local or global minimum of the potential before converging to a constant as $t\rightarrow\infty.$ 
 The scale factor will monotonically increase and $\phi(t)$ will tend towards one of the two minima $\phi(\infty)\rightarrow\pm\lambda$.\footnote{A notable exception to this behavior is the cosmological time crystal \cite{Bains:2015gpv,Easson:2016klq,Easson:2018qgr}. In that scenario, the scalar field converges to a limit cycle in its phase space (instead of a fixed point) despite the presence of Hubble friction. }
 
We can also build an intuition for the gravitational wave speed given those two expectations. For the Einstein frame in 3+1 dimensions with $(\a_,\beta)=(1/2,0),$ the form of the gravitational wave speed (\ref{generalcgw}) is particularly simple and can be written as
\begin{equation}\label{gwspeed2}
c_{gw}^2=1+\a'\Psi(t), \ \ \ \ \ \ \ \ \ \Psi(t) = \frac{\ddot{\phi}-\dot{\phi}(H-2\dot{\phi})}{e^\phi+\a'\dot{\phi}(H-\dot{\phi}/2)}.
\end{equation}
The form of the scalar perturbation speed is significantly more complicated and we will not write it here.  By inspection to (\ref{gwspeed2}), as $t\rightarrow\infty$ we expect $\dot{\phi}\rightarrow 0$ and $\ddot{\phi}\rightarrow 0,$ so that the gravitational wave speed tends to unity. At early times, $\Psi(t)$ will oscillate with initial amplitude of the order $\a'$. 

Recall that the exact solutions for the scale factor had some proportionality to $\a'$. This resulted in the $\a'$ dependence falling out of the field equation for the dilaton such that to determine the value of $\eta$ as in (\ref{generalscalefactor}) and (\ref{c0t0}), we did not need to choose a numerical value for $\a'.$ This is not the case in terms of the numerical integration; we must choose values for $\a'$ and $\lambda,$ as well as the initial values $\phi_0=\phi(0),\dot{\phi}_0=\dot{\phi}(0)$, and $H_0=H(0)$.

\begin{figure}[t]
    \centering
    \begin{minipage}{0.5\textwidth}
        \centering
        \includegraphics[width=0.9\textwidth]{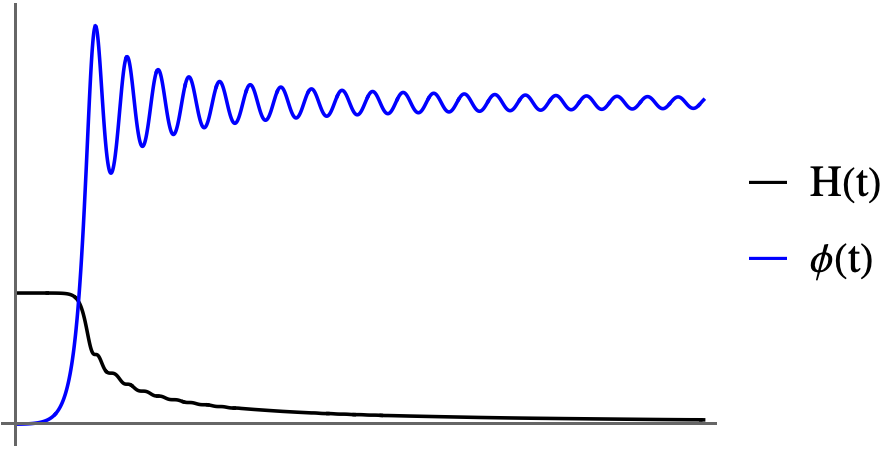} 
    \end{minipage}\hfill
    \begin{minipage}{0.5\textwidth}
        \centering
        \includegraphics[width=0.9\textwidth]{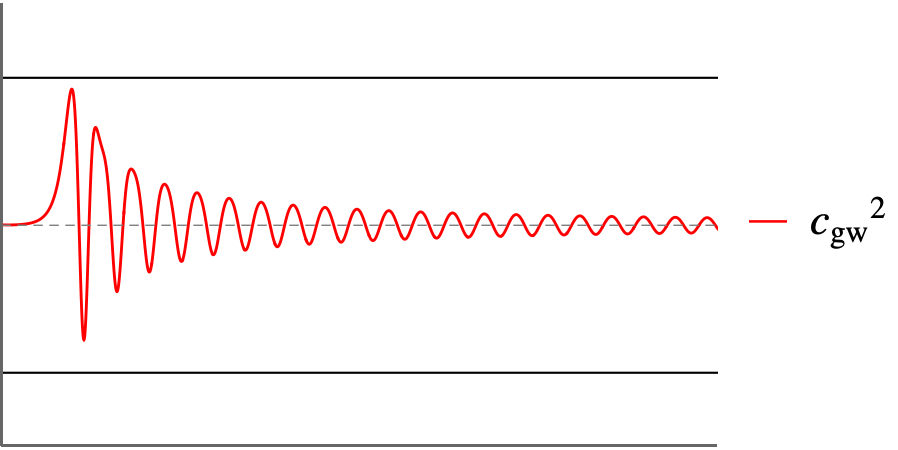} 
    \end{minipage}
    \caption{\textit{Left panel:} The solutions $H(t)$ and $\phi(t)$ for parameter values $\a'=10^{\text{-}5}$, $\lambda=1$, and initial conditions $\phi_0=0.001,\dot{\phi}_0=0,H_0=0.4.$ \textit{Right panel:} The gravitational wave speed associated to the same parameter values and initial conditions. The dashed gray line is $c_{gw}^2=1$ and the two solid black lines are $1\pm O(\a')$. }\label{fig:plots}
\end{figure}

In Figure \ref{fig:plots} we have plotted an example of a stable solution to the string induced model with the potential (\ref{Vhiggs}). As expected, the dilaton oscillates about its local minimum $\phi=\lambda$ and the gravitational wave speed oscillates about unity with initial amplitude of order $\a'.$ The oscillations of both $\phi(t)$ and $c_{gw}^2$ are dampened such that as $t\rightarrow\infty,$ $\phi\rightarrow\lambda$ and $c_{gw}^2\rightarrow 1.$

Not shown are the tensor perturbation quantities $\mathcal{F}_T$ and $\mathcal{G}_T$. Both are non-negative and exhibit damped oscillations about the value $\mathcal{F}_T=\mathcal{G}_T=2$. In addition, the scalar perturbation quantities $\mathcal{F}_S$ and $\mathcal{G}_S$ as in (\ref{FsGs}) evaluated on the solution in Figure \ref{fig:plots} are non-negative and oscillate without any damping. Somewhat surprisingly considering the very different form of each coefficient, $\mathcal{F}_S$ and $\mathcal{G}_S$ are identical such that the scalar sound speed $c_s^2=\mathcal{F}_S/\mathcal{G}_S$ associated with the solution in Figure \ref{fig:plots} is unity at all times. We have therefore provided an example that illustrates that the inclusion of a potential can serve to stabilize the metric perturbations. It is noteworthy that the gravitational wave speed has superluminal and subluminal fluctuations that at most are $O(\a')$ from unity. This is not necessarily problematic in light of the neutron star merger event GW170817 and coincident gamma-ray burst (GRB) GRB 170817A \cite{Monitor:2017mdv}.  Using the time delay between the gravitational wave and GRB observations, $c_{gw}$ is constrained to lie in the range $1-3\times 10^{-15}<c_{gw}/c < 1+7\times 10^{-16}$. Due to the expectation that $\a'\ll 1,$ this bound is not explicitly in conflict with our model.




\section{Discussion} \label{sec:disc}

In this article we studied some of the cosmological solutions to the string induced Galileons uncovered in \cite{Easson:2020bgk}. Specifically, we considered spatially flat FLRW solutions with an isotropic dilaton in different  `frames', corresponding to different choices of the metric ansatz coefficients $\alpha$ and $\beta$. Notably, each frame exhibits cosmic acceleration, except for the Gauss-Bonnet frame for $p=2,...,8$. All of the exact solutions we uncovered -- in the absence of a dilaton potential -- were found to be unstable under scalar or tensor perturbations in $3+1$-dimensions. Focusing particularly on the Einstein frame, we showed including a dilaton potential, \emph{e.g}, the Higgs-like potential, stabilizes the solutions. We stress that, although we chose the Higgs-like potential stabilizes the solution for illustrative purposes, it is straightforward to show other dilaton potentials likewise stabilizes the solution. Thus, the fairly restrictive string induced Galileons, upon including a dilaton potential, admit stable accelerating cosmologies.

There are a number of potentially interesting research avenues worth pursuing. First, the lower dimensional action at tree-level retains the duality symmetry of its higher dimensional counterpart. Upon including $\alpha'$ corrections, the lower-dimensional action no longer has the same symmetry properties of the parent theory. In part this -- aside from the dual frame -- the tree-level and first $\alpha'$ correction are no longer at the same order in a (genus) $g_{s}$ expansion. This is a consequence of the fact we have identified the higher dimensional dilaton $\Phi$ to the lower dimensional scalar field $\phi$, obscuring the connection between $g_{s}$ and $\phi$ post KK reduction. One way around this is to identify the reduced $\Phi$ with some other scalar field $\varphi$, resulting in a theory of multi-field Galileons \cite{Padilla:2010de,Padilla:2010ir,Deffayet:2010zh,Hinterbichler:2010xn}. It would be interesting to pursue this as one could more naturally understand the various $g_{s}$ regimes from the lower dimensional perspective, and more easily incorporate higher genus corrections.

Second, we considered the gravi-dilaton sector of the $\alpha'$-corrected effective string action. More generally, to maintain conformal invariance of the worldsheet string action, we should include the antisymmetric Kalb-Ramond field $B_{\mu\nu}$ with its 3-form gauge curvature $H=dB$.\footnote{Or, we could consider  type IIA or type IIB superstring actions which are known to include additional higher $p$-form gauge fields.} Moreover, in the KK ansatz itself it is natural to introduce Maxwell gauge fields. The resulting dimensionally reduced theory is then a $p+1$-dimensional theory of non-minimally coupled theory of $p$-form covariantized Galileons \cite{Deffayet:2010zh}. 

Other potentially fruitful avenues would be to include additional background matter fields, including the fermionic sectors of the string action, and look for other types of cosmological solutions. Indeed, the tree-level action, via the reduced duality symmetry, admits exact bouncing cosmologies. Moreover, Horndeski theories themselves are known to admit bouncing cosmologies \cite{Qiu:2011cy,Easson:2011zy,Rubakov13-1,Ijjas16-1}.

Let us now briefly discuss how the cosmological solutions uncovered here fair with restrictions imposed by the swampland criteria. Recall the swampland criteria are a set of conjectures \cite{Vafa:2005ui,Ooguri:2006in} (see \cite{Palti:2019pca,vanBeest:2021lhn} for recent reviews), each of which are used to show that not all consistent looking effective field theories (EFTs) can be coupled to gravity in a consistent manner and be UV complete. All EFTs which violate these criteria do not admit a (string theoretic) UV completion, and are said to live in the `swampland' (as opposed to the string landscape). Generalized Horndeski models are a subset of such EFTs, and thus naturally one wonders whether Horndeski gravity belongs to the swampland or string landscape. As laid forth in \cite{Heisenberg:2019qxz,Brahma:2019kch}, subclasses of generalized Galileons, specifically Quintessence, become highly constrained by the de Sitter (dS) conjecture  \cite{Agrawal:2018own,Obied:2018sgi}, where the gradient of a scalar field potential is bounded from below, ruling out metastable de Sitter solutions.\footnote{When we refer to the de Sitter conjecture we really mean the two following conditions: (i) The range $\Delta\phi$ traversed by any scalar field $\phi$ in field theory space is bounded above, $|\Delta\phi|/M_{\text{PL}}\leq \Delta\sim\mathcal{O}(1)$, and (ii) the gradien of the scalar field potential $V(\phi)$ is bounded from below $M_{\text{PL}}|\nabla_{\phi}V|/V>c\sim\mathcal{O}(1)$, where $\Delta,c\in\mathbb{R}^{+}$. Technically condition (i) is known as the range condition while (ii) is the dS constraint, which is the one implying there are no metastable dS solutions emerging in string, ruling out standard $\Lambda$CDM cosmology. Condition (ii) also implies models with the largest possible $V'/V$ satisfying current observation constraints on dark energy imposes $V(\phi)=V_{0}e^{\lambda\phi}$, for $\lambda\approx.6$ (see, \emph{e.g.}, \cite{Agrawal:2018own}).} Nonetheless, subclasses of generalized Galileons (even those beyond Quintessence) exist within the habitable landscape.

More recently, a no-go theorem reveals that theories of gravity with extra dimensions satisfying reasonable assumptions do not generally admit solutions with cosmic acceleration without violating the null energy condition in the compactified space \cite{Montefalcone:2020vlu}. The no-go theorem is based on two sets of constraints: (1) metric based constraints and (2) swampland criteria, particularly the dS constraint. The metric based constraints assume the higher dimensional parent theory and the compactified theory are each described by a respective Einstein-Hilbert action; the higher dimensional metric is (conformally) Ricci flat in its compactified directions (which ansatz (\ref{metansatz2}) with $\phi=\phi(t)$ is a special case of), and the $p+1$-dimensional geometry is spatially flat FLRW. Together, constraints (1) and (2) imply cosmic acceleration is impossible unless the NEC is violated in the compactified space. That is, for $T_{MN}$ being the energy-momentum tensor of the parent theory (in Einstein frame) such that $p_{n}=T^{n}_{\;n}$ is the `pressure' of the $n$-dimensional compactified stress tensor and $\rho(t)$ is the higher dimensional energy density $T_{tt}$, then $\rho+p_{n}<0$ for some time. Quintessence is again a favored model capable of satisfying both (1) and (2), and can be used to model dark energy within observational bounds \cite{Montefalcone:2020vlu}.  

While the low energy effective string action considered here, at tree level, seemingly must violate the NEC in the compactified space to admit realistic solutions of cosmic acceleration, the $\alpha'$-corrected action may evade this restriction. This is simply because the higher dimensional metric ansatz is assumed to be a solution to the $\alpha'$-model including the Gauss-Bonnet term, such that it is possible to circumvent assumption (1), such that the NEC may not need to be violated in the compactified directions. It would be interesting to study this in more detail and see whether the no-go theorem of \cite{Montefalcone:2020vlu} may be generalized to incorporate models of the type explored here, which may provide additional constraints on string induced Galileons. We hope to return to this question, as well as the aforementioned directions, in the near future.

\noindent\section*{ACKNOWLEDGMENTS}

We are pleased to thank Eoin O Colgain for helpful correspondence. DE is partially supported by a grant from FQXi. AS is supported by the Simons Foundation \emph{It from Qubit} collaboration.

\appendix

\section{Low energy effective string actions}\label{sec:apploweeffact}

We now comment on the different forms of the bosonic sector of the low-energy effective string actions utlized in this work. For additional pedagogical details, see \emph{e.g.} \cite{Gasperini07-1}. For convenience we drop hatted notation indicating we are in $D$-dimensions. 

The $D$-dimensional low-energy effective string action arises from demanding the two-dimensional Polyakov string action maintain conformal symmetry as the model is quantized at each order in a quantum loop expansion. The quantum loop expansion corresponds to an expansion in $\alpha'=\ell_{s}^{2}$, the string length. At tree level, the (gravi-dilaton) $D$-dimensional low energy effective action is 
\beq I_{\text{tree}}=-\frac{1}{2\lambda^{D-2}_{s}}\int d^{D}X\sqrt{-g}e^{-\Phi}[R+(\nabla\Phi)^{2}]\;.\eeq
The parameter $\alpha'$ provides a natural length scale for the model such that when the background curvature is large with respect to $\alpha'$ correspond to small curvatures in string units, and in this way an expansion in $\alpha'$ corresponds to a higher derivative expansion of the metric. As curvature grows, the $\alpha'$ corrections to the tree level action become increasingly important.

Unlike the tree-level action, it turns out higher order $\alpha'$ corrections to the Polyakov string action are not uniquely fixed due to ambiguities in field redefinitions. Correspondingly, there are many possible higher derivative $D$-dimensional actions which are allowed. The general allowed form of the gravi-dilaton sector of the $D$-dimensional action is 
\beq 
\begin{split}
I_{\alpha'}&=\frac{\alpha'}{2\lambda^{D-2}_{s}}\int d^{D}X\sqrt{-g}e^{-\Phi}a'_{0}\biggr\{R^{2}_{\mu\nu\alpha\beta}+a'_{1}R^{2}_{\mu\nu}+a'_{2}R^{2}+a'_{3}R^{\mu\nu}(\nabla_{\mu}\Phi)(\nabla_{\nu}\Phi)\\
&+a'_{4}R(\nabla\Phi)^{2}+a'_{5}R\Box\Phi+a'_{6}(\Box\Phi)^{2}+a'_{7}\Box\Phi(\nabla\Phi)^{2}+a'_{8}(\nabla\Phi)^{4}\biggr\}\;.
\end{split}
\eeq 
Through an appropriate transformation of the metric and dilaton \cite{Metsaev87-1}, at most seven of the nine coefficients $\{a'_{i}\}$ can be fixed, however, generally leaving multiple low energy effective actions compatible with string quantization at $\alpha'$. A central action of interest to us is (\ref{stringact1})
\beq I_{1}=-\frac{1}{2\lambda_{s}^{D-2}}\int d^{D}X\sqrt{-g}e^{-\Phi}\left[R+(\nabla\Phi)^{2}-\frac{\alpha'}{4}\mathcal{L}_{\text{GB}}+\frac{\alpha'}{4}(\nabla\Phi)^{2}\right]\;.\eeq
The Gauss-Bonnet term was also found to arise in the low-energy effective action for heterotic string theory \cite{Zwiebach:1985uq} and keeps the theory free of ghost instabilities. 

Another action which exhibits covariantized Galileons and includes a Gauss-Bonnet term is\footnote{Action $I_{1}$ follows from the field redefinitions $g_{\mu\nu}\to g_{\mu\nu}+4\alpha'[R_{\mu\nu}-\nabla_{\mu}\Phi\nabla_{\nu}\Phi+g_{\mu\nu}(\nabla\Phi)^{2}]$ and $\Phi\to\Phi+\alpha'[R+(2(D-1)-3)(\nabla\Phi)^{2}]$, while $I_{2}$ is obtained by performing the field redefinition $g_{\mu\nu}\to g_{\mu\nu}+4\alpha'R_{\mu\nu}$ and $\Phi\to\Phi+\alpha'[R-(\nabla\Phi)^{4}]$.}
\beq 
\begin{split}
I_{2}&=-\frac{1}{2\lambda_{s}^{D-2}}\int d^{D}X\sqrt{-g}e^{-\Phi}\biggr[R+(\nabla\Phi)^{2}\\
&-\frac{\alpha'}{4}[\mathcal{L}_{\text{GB}}-4G^{\mu\nu}(\nabla_{\mu}\Phi)(\nabla_{\nu}\Phi)+2\Box\Phi(\nabla\Phi)^{2}-(\nabla\Phi)^{4}\biggr]\;.
\end{split}
\eeq
We won't explore this model here, however, note that had we started from this action instead and performed the same KK reduction, the only difference from the $p+1$-dimensional action we considered are constant shifts to the coefficients $a_{i}$.


\section{Numerical values of expansion rates}\label{AppendixForRates}

We will now illustrate the process of obtaining all relevant numerical values associated to a complete solution in the various frames.

\subsection{Einstein frame}
The Einstein frame conditions $f_{E}=0$ and $a_{1}=-\frac{1}{2}$ result in the two pairs $(\a_{\pm},\beta_{\pm})$ as in (\ref{abplus}) and (\ref{abminus}). We will find that solutions associated with $(\alpha_+,\beta_+)$ are surprisingly different than those associated with $(\alpha_-,\beta_-)$, despite these solutions arising from the same parent theory.  The Friedmann equation is 
\begin{equation}\label{Fried2}
0=\mRd^2\mR^{p-2}\Bigg[\bP_0(q)-\alpha'\bP_1(q)\Bigg( \frac{\mRd}{\mR^{1+\alpha q}}\Bigg)^2\Bigg],
\end{equation}
where $\bP_0(q)$ and $\bP_1(q)$ are the polynomials
\begin{equation}\label{P0}
\bP_0(q) = p(p-1)-\frac{1}{2}q^2,
\end{equation}
\begin{equation}\label{P1}
\bP_1(q) =\frac{1}{4}\Big[ p(p-1)(p-2)(p-3)-8\alpha qp(p-1)(p-2)-\frac{3}{2}q^2a_2 p(p-1)-2q^3 a_3p-q^4(2\alpha a_3+3a_4)\Big].
\end{equation}
The equation of motion for the dilaton, (\ref{phi3}), is
\begin{equation}\label{phi4}
\begin{split}
0=\mR^p\Bigg[\bA\Big(\frac{\mRd}{\mR}\Big)^2+\bB\frac{\mRdd}{\mR}-\a'\mR^{-2\a q}\Bigg(\bC\Big(\frac{\mRd}{\mR}\Big)^4-\bD\Big(\frac{\mRd}{\mR}\Big)^2\Big(\frac{\mRdd}{\mR}\Big)\Bigg)\Bigg],
\end{split}
\end{equation}
where the coefficients simplify to
\begin{equation}
\begin{split}
\bA&=q(1-p), \ \ \ \ \ \ \ \ \ \bB=-q, \\
\bC&=\frac{1}{4}\Big[-p(p-1)(p-2)\big(2\a(p-3)+a_2q)+2q^3(2\a a_3+3a_4)(\a q+2) \\
& \ \ \ \ \ \ \ \ \ +qp(p-1)\big(a_2+q(\a a_2-2a_3)\big)+4pq^2(a_3-a_4 q)\Big], \\
\bD&=\frac{1}{4}\Big[2q^2\big(pa_3+2q(2\a a_3+3a_4)\big)+p(p-1)\big(8\a(p-2)+a_2 q\big)+2pq\big(a_2(p-1)+2a_3 q\big) \Big].
\end{split}
\end{equation}
The coefficients $a_2,$ $a_3$, and $a_4$ are given in (\ref{ais}) and are also determined solely by the choice of internal dimension $n$ and bulk spatial dimension $p$, since $\alpha$ and $\beta$ are to be set to either (\ref{abplus}) or (\ref{abminus}).

\subsection*{3+1 dimensional solution}

When $p=3$ and $n=6,$ the constants $(\a_{\pm},\beta_{\pm})$ from (\ref{abplus}), (\ref{abminus}) are
\begin{equation}
(\a_+, \beta_+) = \Big(\frac{1}{2},0\Big), \ \ \ \ \ \ \ \ \ \ (\a_-,\beta_-)=\Big(-\frac{1}{4},\frac{1}{4}\Big),
\end{equation}
and for completeness, the coefficients (\ref{ais}) are  
\begin{equation}\label{p3ais}
\begin{split}
(\alpha_+,\beta_+): \ \ \ & \ \ \ (a_2,a_3,a_4) = (4,3,-1), \\
(\alpha_-,\beta_-): \ \ \ & \ \ \ (a_2,a_3,a_4)=(11/2, 15/4, -19/16).
\end{split}
\end{equation}
Inserting the values for $(a_+,\beta_+),$ the polynomial (\ref{EinsteinPee}) becomes
\begin{equation}\label{P36plus}
\mathcal{P}^{+}_{3,6}(q) = \frac{3}{8}\Big(q^5+22q^4+112 q^3+352 q^2+336 q+96\Big),
\end{equation}
while for $(\a_-,\beta_-),$ 
\begin{equation}\label{P36minus}
\mathcal{P}^{-}_{3,6}(q) = -\frac{51}{32}q^5+\frac{735}{64}q^4-\frac{345}{16}q^3+69q^2+153q-18.
\end{equation}
Both (\ref{P36plus}) and (\ref{P36minus}) have three real roots and two complex roots. The real roots are 
\begin{equation}
\mathcal{P}^{+}_{3,6}(q^*)=0 \ \ \ \Rightarrow \ \ \ q_a^* \approx -16.41, \ \ \ q_b^*\approx -0.701, \ \ \ q_c^*\approx -0.584,
\end{equation}
\begin{equation}
\mathcal{P}^{-}_{3,6}(q^*) = 0 \ \ \ \Rightarrow \ \ \ q_d^* \approx-1.351, \ \ \ q_e^*\approx0.112, \ \ \ q_f^*\approx 6.493.
\end{equation}
With the real roots in hand, we now need to check the second reality condition, II of (\ref{realcondish}), which is $\frac{\bar{P}_0(q^*)}{\bar{P}_1(q^*)}>0.$ For the roots to $\mathcal{P}^{+}_{3,6},$ we find
\begin{equation}
\frac{\bP_0(q_a^*)}{\bP_1(q_a^*)}\approx -0.01, \ \ \ \ \ \frac{\bP_0(q_b^*)}{\bP_1(q_b^*)} \approx 4.31, \ \ \ \ \ \ \frac{\bP_0(q_c^*)}{\bP_1(q_c^*)}\approx 4.38,
\end{equation}
which illustrates we need to toss out $q_a^*$. For the three real roots to $\mathcal{P}^{-}_{3,6},$ the reality condition is
\begin{equation}
\frac{\bP_0(q_d^*)}{\bP_1(q_d^*)}\approx -0.62, \ \ \ \ \ \ \frac{\bP_0(q_e^*)}{\bP_1(q_e^*)}\approx 34.6, \ \ \ \ \ \ \frac{\bP_0(q_f^*)}{\bP_1(q_f^*)}\approx -0.04,
\end{equation}
therefore, we discard $q_d^*$ and $q_f^*$. Finally, recalling that the expansion rate $\eta=-\frac{1}{\alpha q}$, we can obtain the expansion powers for each root. For $\eta=-\frac{1}{\a_+  q^*}$ and the valid roots $q_b^*$ and $q_c^*$, we have
\begin{equation}
\eta \approx 2.853, \ 3.422,
\end{equation}
respectively, while for $\eta=-\frac{1}{\a_- q^*},$ the one valid root $q_e^*$ results in 
\begin{equation}
\eta \approx 35.66.
\end{equation}
It is straightforward to work out the expansion rates for the other dimensions using the same approach. The results are presented in Section \ref{sec:FRLWcosmo} in Tables \ref{plussolstable} and \ref{minussolstable2}.

\subsection{String frame}

The string frame is defined by setting $f_E=-1$ and $a_1 = 1,$ which fixes the coefficients $\a$ and $\beta$ to be
\begin{equation}\label{abstring}
\a=-\frac{2n}{(p-1)(p+n-1)}, \ \ \ \ \ \ \ \beta = \frac{2}{p+n-1}.
\end{equation}
The structure of the solution associated to this choice of parameters is identical to  (\ref{generalscalefactor}) and (\ref{c0t0}), however the polynomials, and therefore roots $q^*$, are not the same. Mainly, the polynomial (\ref{EinsteinPee}) is modified to be
\begin{equation}\label{stringpoly}
\mathcal{P}'_{p,n}(q)=P'_1(q)\big(A'+B'(1+\a q)\big)-P_0'(q)\big( C'-D'(1+\a q)\big),
\end{equation}
where we now have
\begin{equation}
\begin{split}
P_0'(q)&=p(p-1)-2pq+q^2, \\
P_1'(q)&=\frac{1}{4}\Big[p(p-1)(p-2)(p-3)-4pq(p-1)(p-2)(1+2\a)-\frac{3}{2}p(p-1)a_2 q^2 \\
& \ \ \ \ \ \ -2pa_3q^3-q^4\big((1+2\a)a_3+3a_4\big)\Big], \\
A'&=-p(p-1)-q^2-q(1-2p), \\
B'&=2(q-p), \\
C'&=\frac{1}{4}\Big[-p(p-1)(p-2)\big((p-3)(1+2\a)+a_2 q)+q^3\big((1+2\a)a_3+3a_4\big)\big(q(1+2\a)+4\big) \\
& \ \ \ \ \ \ +\frac{1}{2}pq(p-1)\Big(q\big( (1+2\a)a_2-4a_3\big)+2a_2\Big)-4pq^2(a_4 q-a_3)\Big], \\
D'&= \frac{1}{4}\Big[q^2\Big(2pa_3+\big( (1+2\a)a_3+3a_4\big) \Big)+p(p-1)\big(4(p-2)(1+2\a)+2a_2  q\big) \\
& \ \ \ \ \ \ +2pq\big(a_2(p-1)+2a_3 q\big)\Big].
\end{split}
\end{equation}
The coefficients $a_2, a_3$ and $a_4$ are again given by (\ref{ais}) with $\a$ and $\beta$ as in (\ref{abstring}). For a given root to (\ref{stringpoly}) $q^*$, the reality conditions are the same as before,
\begin{equation}
\text{I:} \ \ \ q^*\in\mathbb{R}, \ \ \ \ \ \ \ \ \text{II:} \ \ \ \frac{P_0'(q^*)}{P_1'(q^*)}>0,
\end{equation}
and the expansion rate is again $\eta = -\frac{1}{\a q^*}.$ 

\subsubsection*{3+1 dimensional solution}

When $p=3$ and $n=6,$ $\a$ and $\beta$ become
\begin{equation}
(\a,\beta) = \Big(-\frac{3}{4},\frac{1}{4}\Big)
\end{equation}
while the remaining $a_i$ coefficients are
\begin{equation}
(a_2,a_3,a_4) = \Big(\frac{15}{2},-6,-\frac{31}{16}\Big). 
\end{equation}
The polynomial (\ref{stringpoly}) becomes
\begin{equation}
\mathcal{P}'_{3,6}(q)=-\frac{3}{128}\Big(479 q^5-114q^4-5416q^3+12032 q^2-8448q+768\Big),
\end{equation}
which also has three real and two complex roots. The three real roots are
\begin{equation}
\mathcal{P}_{3,6}'(q^*)=0 \ \ \ \Rightarrow \ \ \ q^*_a\approx -4.17, \ \ \ q^*_b\approx 0.106, \ \ \ q^*_c\approx 1.53,
\end{equation}
and the respective reality conditions read
\begin{equation}
\frac{P_0'(q^*_a)}{P_1'(q^*_a)}\approx -0.06, \ \ \ \ \ \ \frac{P_0'(q^*_b)}{P_1'(q^*_b)}\approx 38.6, \ \ \ \ \ \ \ \frac{P_0'(q^*_c)}{P_1'(q^*_c)}\approx -0.72,
\end{equation}
therefore we discard $q^*_a$ and $q^*_c.$ For $q^*_b,$ the expansion rate is 
\begin{equation}
\eta \approx 12.55.
\end{equation}
Table \ref{stringtable} in Section \ref{sec:FRLWcosmo} summarizes the results for $p=2,...,8.$

\subsection{Dual frame}\label{appen:dualframe}

Here, we will briefly go through the exercise of obtaining as much information as possible about the Hubble constant as in (\ref{ADS1}) for two particular values of $\beta$. Consider the 6-dimensional ($p=5$) case. Setting $\beta=+1$ results in the field equation for the dilaton (\ref{tildeBigP}) reducing to 
\begin{equation}\label{betap1}
\tilde{\mathcal{P}}^{\beta=1}_{5,4}(q)=\frac{5}{2} \left(386 q^5+2493 q^4+4560 q^3+4200 q^2+640 q-1080\right),
\end{equation}
while setting $\beta=-1$ yields
\begin{equation}\label{betam1}
\tilde{\mathcal{P}}^{\beta=\text{-}1}_{5,4}(q)=\frac{5}{2} \left(1986 q^5-5915 q^4+4560 q^3+1320 q^2-4480 q+1800\right).
\end{equation}
There are three real roots to (\ref{betap1}), which are
\begin{equation}
q_1^*\approx-4.25, \ \ \ \ q_2^* \approx -0.987, \ \ \ \  q_3^*\approx 0.368,
\end{equation}
as well as three real roots to (\ref{betam1}):
\begin{equation}
q_4^*\approx -0.841, \ \ \ \ q_5^*\approx 0.578, \ \ \ \ q_6^*\approx 1.80.
\end{equation}
Checking the reality condition for the $\beta=+1$ case shows
\begin{equation}
\frac{\tilde{P}_0(q_2^*)}{\tilde{P}_1(q_2^*)}<\frac{\tilde{P}_0(q_1^*)}{\tilde{P}_1(q_1^*)} < 0, \ \ \ \ \ \ \ \  \frac{\tilde{P}_0(q_3^*)}{\tilde{P}_1(q_3^*)} >0,
\end{equation}
while for the $\beta=-1$ case,
\begin{equation}
\frac{\tilde{P}_0(q_5^*)}{\tilde{P}_1(q_5^*)}<\frac{\tilde{P}_0(q_6^*)}{\tilde{P}_1(q_6^*)}<0, \ \ \ \ \ \ \ \ \ \frac{\tilde{P}_0(q_4^*)}{\tilde{P}_1(q_4^*)}>0.
\end{equation}
We conclude that there is one solution to the field equations in $p=5$ for $\beta=+1$ and one solution for $\beta=-1.$ Defining the shorthand for the Hubble constant in (\ref{ADS1}) as
\begin{equation}
H_0\equiv \epsilon \frac{\bar{h}}{\ell_s},
\end{equation}
where the dimensionless quantity in the numerator is
\begin{equation}
\bar{h}=\sqrt{\frac{\tilde{P}_0(q^*)}{\tilde{P}_1(q^*)}},
\end{equation}
we find that
\begin{equation}
\bar{h}^{\beta=1}\approx 0.539, \ \ \ \ \ \ \ \ \bar{h}^{\beta=\text{-}1}\approx0.292.
\end{equation}
It is straightforward to repeat this exercise for $p=2,...,8$. The results are summarized in Tables \ref{betaplustable} and \ref{betaminustable}.

\begin{table}[H]
\begin{center}
\begin{tabular}{c| c | c | c }
$p,n$  & Real roots & Valid roots & $\bar{h}$ \\ \hline 
2, 7 & 2 & 1 & 0.757 \\
3, 6 & 3 & 1 & 0.669  \\
4, 5 & 3 & 1 & 0.599  \\
5, 4 &  3& 1 & 0.539 \\
6, 3 & 3 & 1 & 0.482  \\
7, 2 & 2 &1 & 0.426  \\
8, 1 & 2 & 0 & $\sim$ 
\end{tabular}
\end{center}
\caption{Solutions associated to $\beta=+1.$ Note that we do not consider any trivial roots $q^*=0$ in this table. Interestignly, the polynomial (\ref{tildeBigP}) is quartic when $p=7$ and when $p=8$, the polynomial has only odd powers of $q$ and no roots satisfy (\ref{realC2}). In $p=2$, the polynomial has no $q^0$ term, but otherwise cases $p=3,4,5,6$ produce polynomials with similar structure. }
\label{betaplustable}
\end{table}

\begin{table}[H]
\begin{center}
\begin{tabular}{c| c | c | c }
$p,n$  & Real roots & Valid roots & $\bar{h}$ \\ \hline 
2, 7 & 2 & 1 & 0.294 \\
3, 6 & 3 & 1 & 0.294  \\
4, 5 & 3 & 1 & 0.293  \\
5, 4 &  3& 1 & 0.292 \\
6, 3 & 3 & 1 & 0.291  \\
7, 2 & 3 &1 & 0.289 \\
8, 1 & 3 & 2 & 0.286, 0.094 
\end{tabular}
\end{center}
\caption{Solutions associated to $\beta=-1.$ Note that we do not consider any trivial roots $q^*=0$ in this table. Surprisingly, $\bar{h}$ is nearly the same value in all dimensions (with the exception of the second valid root in $p=8$), suggesting a type of universality class. In fact, this observation is consistent for a wide range of negative values of $\beta$. Note, however, this universality is not present for $-1<\beta<0;$ for that range, the $\bar{h}$ values vary more similarly to Table \ref{betaplustable}.}
\label{betaminustable}
\end{table}


\section{Linear perturbations to FLRW} \label{app:linperts}


 Here we provide a very brief review of analysis of linear perturbations about a FLRW background in a Horndeski action, first carried out in \cite{DeFelice:2011bh} (for a recent review, see \cite{Kobayashi:2019hrl}).

Consider a spatially flat FLRW background metric in ADM form with a homogeneous scalar field,
\beq ds^{2}=-N^{2}(t)dt^{2}+\gamma_{ij}(dx^{i}+N^{i}dt)(dx^{j}+N^{j}dt)\;,\quad \phi=\phi(t)\;.\eeq
The action for the generalized Galileons (\ref{Horndaction}) will then generically be a function of $N$ and its first derivative, and $\mR$ and $\phi$ and their first two derivatives. Here we specifically consider scalar and tensor perturbations about this background.\footnote{General linear perturbations about an FLRW background may be decomposed into scalar, tensor and vector modes, however, the vector components are generally uninteresting since they are non-dynamical in general relativity and scalar-tensor theories. We therefore neglect such contributions as typically done.} Via general covariance some perturbation variables may be eliminated by a gauge transformation. In particular, we use the unitary gauge such that $\phi\to\phi+\delta\phi$ with $\delta\phi=0$, such that all fluctuations are in the metric 
\beq N=1+\delta N\;,\quad N_{i}=\partial_{i}\psi\;,\quad \gamma_{ij}=\mathcal{R}^{2}e^{2\zeta}(e^{h})_{ij}\;,\eeq
with
\beq (e^{h})_{ij}=\delta_{ij}+h_{ij}+\frac{1}{2}h_{ik}h_{kj}+...\;.\eeq
Here $\delta N$, $\psi$ and $\zeta$ are scalar perturbations while $h_{ij}$ are tensor perturbations, \emph{i.e.}, gravitational waves, obeying transverse traceless conditions: $\partial^{i}h_{ij}=h^{i}_{\;i}=0$. Substituting the FLRW solution into the action (\ref{Horndaction}) and expanding up to second order in perturbations, one finds the second order action to be given by $ I^{(2)}=I^{(2)}_{\text{S}}+I^{(2)}_{T}$, where the scalar and tensor actions quadratic in perturbations, $I_{S}^{(2)}$ and $I_{T}^{(2)}$, respectively, are \cite{Kobayashi:2011nu,Kobayashi:2019hrl}
\beq 
\begin{split}
I^{(2)}_{S}&=\int dt d^{3}x\mathcal{R}^{3}\biggr[-3\mathcal{G}_{T}\dot{\zeta}^{2}+\frac{\mathcal{F}_{T}}{\mathcal{R}^{2}}(\partial\zeta)^{2}+\Sigma\delta N^{2}-2\Theta\delta N\frac{\partial^{2}\psi}{\mathcal{R}^{2}}+2\mathcal{G}_{T}\dot{\zeta}\frac{\partial^{2}\psi}{\mathcal{R}^{2}}\\
&+6\Theta\delta N\dot{\zeta}-2\mathcal{G}_{T}\delta N\frac{\partial^{2}\zeta}{\mathcal{R}^{2}}\biggr]\;,
\end{split}
\eeq
and
\beq I^{(2)}_{T}=\frac{1}{8}\int dtd^{3}x\mathcal{R}^{3}\left[\mathcal{G}_{T}\dot{h}_{ij}^{2}-\frac{\mathcal{F}_{T}}{\mathcal{R}^{2}}(\partial_{k}h_{ij})^{2}\right]\;.\eeq
The coefficients $\mathcal{G}_{T},\mathcal{F}_{T},\Sigma$ and $\Theta$ depend on the form functions appearing in the Horndeski action. Explicitly, 
\beq \mathcal{G}_{T}\equiv2\left[G_{4}-2XG_{4,X}-X(H\dot{\phi}G_{5,X}-G_{5,\phi})\right]\;,\label{eq:GTgen}\eeq
\beq \mathcal{F}_{T}\equiv2\left[G_{4}-X\left(\ddot{\phi}G_{5,X}+G_{5,\phi}\right)\right]\;,\label{eq:FTgen}\eeq

\beq\label{Sigma}
\begin{split}
\Sigma&\equiv XK_{X}+2X^{2}K_{XX}+12 H\dot{\phi}XG_{3,X}+6H\dot{\phi}X^{2}G_{3,XX}-2XG_{3,\phi}-2X^{2}G_{3,\phi X}\\
&-6H^{2}G_{4}+6\biggr(H^{2}[7XG_{4,X}+16 X^{2}G_{4,XX}+4X^{3}G_{4,XXX}]\\
&-H\dot{\phi}(G_{4,\phi}+5XG_{4,\phi X}+2X^{2}G_{4,\phi XX})\biggr)\\
&+30 H^{3}\dot{\phi} XG_{5,X}+26 H^{3}\dot{\phi}X^{2}G_{5,XX}+4H^{3}\dot{\phi}X^{3}G_{5,XXX}\\
&-6H^{2}X(6G_{5,\phi}+9XG_{5,\phi X}+2X^{2}G_{5,\phi XX})\;,
\end{split}
\eeq

\beq 
\begin{split}
\Theta&\equiv -\dot{\phi}XG_{3,X}+2HG_{4}-8HX G_{4,X}-8H X^{2}G_{4,XX}+\dot{\phi}G_{4,\phi}+2X\dot{\phi}G_{4,\phi X}\\
&-H^{2}\dot{\phi}(5X G_{5,X}+2X^{2} G_{5,XX})+2HX(3G_{5,\phi}+2XG_{5,\phi X})\;.
\end{split}
\eeq

Since time derivatives of $\delta N$ and $\psi$ do not appear in the quadratic action for the scalar perturbation, their equations of motion impose constraint equations
\beq \Sigma\delta N-\Theta\frac{1}{\mathcal{R}^{2}}\partial^{2}\psi+3\Theta\zeta-\mathcal{G}_{T}\frac{1}{\mathcal{R}^{2}}\partial^{2}\zeta=0\;,\eeq
\beq \Theta\delta N-\mathcal{G}_{T}\dot{\zeta}=0\;.\eeq
Using these constraints to replace $\delta N$ and $\psi$ in terms of $\zeta$ in the scalar action $I^{(2)}_{S}$, one now has
\beq I^{(2)}_{S}=\int dt d^{3}x\mathcal{R}^{3}\left[\mathcal{G}_{S}\dot{\zeta}^{2}-\frac{\mathcal{F}_{S}}{\mathcal{R}^{2}}(\partial\zeta)^{2}\right]\;,\eeq
with
\beq \label{FsGs}
 \mathcal{G}_{S}\equiv\frac{\Sigma}{\Theta^{2}}\mathcal{G}^{2}_{T}+3\mathcal{G}_{T}\;,\quad \mathcal{F}_{S}=\frac{1}{\mathcal{R}}\frac{d}{dt}\left(\frac{\mathcal{R}}{\Theta}\mathcal{G}_{T}^{2}\right)-\mathcal{F}_{T}\;.\eeq

The propagation speeds for the scalar and tensor modes $c_{s}^{2}$ and $c^{2}_{gw}$, respectively, are given by 
\beq c_{s}^{2}\equiv\frac{\mathcal{F}_{S}}{\mathcal{G}_{S}}\;,\quad c_{gw}^2 \equiv\frac{\mathcal{F}_{T}}{\mathcal{G}_{T}}\;.\label{eq:speedsgen}\eeq
These must be positive, $c_{gw}^{2},c_{s}^{2}>0$, as otherwise the perturbations will exhibit exponential growth leading to a gradient instability. Moreover, to guarantee no ghost instabilities (namely, insisting on positivity of the kinetic terms for $h_{ij}$ and $\zeta$), one requires $\mathcal{G}_{T},\mathcal{G}_{S}>0$. Altogether, the stability conditions of a given cosmological model are summarized by 
\beq \mathcal{G}_{T}>0\;,\quad \mathcal{F}_{T}>0\;,\quad \mathcal{G}_{S}>0\;,\quad \mathcal{F}_{S}>0\;.\eeq
One often further imposes the subluminality conditions, such that the speeds (\ref{eq:speedsgen}) are less than or equal to one,
\beq c_{gw}^{2}\leq1\;,\quad c_{s}^{2}\leq1\;,\eeq
so as to avoid potential causal paradoxes and also ensure the effective theory is UV complete in the standard Wilsonian sense.

\bibliography{referencesGR}

\providecommand{\href}[2]{#2}\begingroup\raggedright\begin{thebibliography}{10}

\bibitem{Clifton:2011jh}
T.~Clifton, P.~G. Ferreira, A.~Padilla and C.~Skordis, \emph{{Modified Gravity
  and Cosmology}},
  \href{http://dx.doi.org/10.1016/j.physrep.2012.01.001}{\emph{Phys. Rept.}
  {\bf 513} (2012) 1--189}, [\href{https://arxiv.org/abs/1106.2476}{{\tt
  1106.2476}}].

\bibitem{Lovelock:1971yv}
D.~Lovelock, \emph{{The Einstein tensor and its generalizations}},
  \href{http://dx.doi.org/10.1063/1.1665613}{\emph{J. Math. Phys.} {\bf 12}
  (1971) 498--501}.

\bibitem{Lovelock:1972vz}
D.~Lovelock, \emph{{The four-dimensionality of space and the einstein tensor}},
  \href{http://dx.doi.org/10.1063/1.1666069}{\emph{J. Math. Phys.} {\bf 13}
  (1972) 874--876}.

\bibitem{Horndeski74-1}
G.~W. Horndeski, \emph{{Second-order scalar-tensor field equations in a
  four-dimensional space}},
  \href{http://dx.doi.org/10.1007/BF01807638}{\emph{Int. J. Theor. Phys.} {\bf
  10} (1974) 363--384}.

\bibitem{Glavan:2019inb}
D.~z. Glavan and C.~Lin, \emph{{Einstein-Gauss-Bonnet gravity in 4-dimensional
  space-time}},
  \href{http://dx.doi.org/10.1103/PhysRevLett.124.081301}{\emph{Phys.\ Rev.\
  Lett.} {\bf 124} (2020) 081301},
  [\href{https://arxiv.org/abs/1905.03601}{{\tt 1905.03601}}].

\bibitem{Deffayet:2009wt}
C.~Deffayet, G.~Esposito-Farese and A.~Vikman, \emph{{Covariant Galileon}},
  \href{http://dx.doi.org/10.1103/PhysRevD.79.084003}{\emph{Phys. Rev. D} {\bf
  79} (2009) 084003}, [\href{https://arxiv.org/abs/0901.1314}{{\tt
  0901.1314}}].

\bibitem{Deffayet:2009mn}
C.~Deffayet, S.~Deser and G.~Esposito-Farese, \emph{{Generalized Galileons: All
  scalar models whose curved background extensions maintain second-order field
  equations and stress-tensors}},
  \href{http://dx.doi.org/10.1103/PhysRevD.80.064015}{\emph{Phys. Rev. D} {\bf
  80} (2009) 064015}, [\href{https://arxiv.org/abs/0906.1967}{{\tt
  0906.1967}}].

\bibitem{Deffayet:2011gz}
C.~Deffayet, X.~Gao, D.~Steer and G.~Zahariade, \emph{{From k-essence to
  generalised Galileons}},
  \href{http://dx.doi.org/10.1103/PhysRevD.84.064039}{\emph{Phys. Rev. D} {\bf
  84} (2011) 064039}, [\href{https://arxiv.org/abs/1103.3260}{{\tt
  1103.3260}}].

\bibitem{Kobayashi:2011nu}
T.~Kobayashi, M.~Yamaguchi and J.~Yokoyama, \emph{{Generalized G-inflation:
  Inflation with the most general second-order field equations}},
  \href{http://dx.doi.org/10.1143/PTP.126.511}{\emph{Prog. Theor. Phys.} {\bf
  126} (2011) 511--529}, [\href{https://arxiv.org/abs/1105.5723}{{\tt
  1105.5723}}].

\bibitem{Charmousis14-1}
C.~Charmousis, \emph{{From Lovelock to Horndeski`s Generalized Scalar Tensor
  Theory}}, \href{http://dx.doi.org/10.1007/978-3-319-10070-8_2}{\emph{Lect.
  Notes Phys.} {\bf 892} (2015) 25--56},
  [\href{https://arxiv.org/abs/1405.1612}{{\tt 1405.1612}}].

\bibitem{Nicolis:2008in}
A.~Nicolis, R.~Rattazzi and E.~Trincherini, \emph{{The Galileon as a local
  modification of gravity}},
  \href{http://dx.doi.org/10.1103/PhysRevD.79.064036}{\emph{Phys. Rev. D} {\bf
  79} (2009) 064036}, [\href{https://arxiv.org/abs/0811.2197}{{\tt
  0811.2197}}].

\bibitem{Silva:2009km}
F.~P. Silva and K.~Koyama, \emph{{Self-Accelerating Universe in Galileon
  Cosmology}}, \href{http://dx.doi.org/10.1103/PhysRevD.80.121301}{\emph{Phys.
  Rev. D} {\bf 80} (2009) 121301}, [\href{https://arxiv.org/abs/0909.4538}{{\tt
  0909.4538}}].

\bibitem{Creminelli10-1}
P.~Creminelli, A.~Nicolis and E.~Trincherini, \emph{{Galilean Genesis: An
  Alternative to inflation}},
  \href{http://dx.doi.org/10.1088/1475-7516/2010/11/021}{\emph{JCAP} {\bf 1011}
  (2010) 021}, [\href{https://arxiv.org/abs/1007.0027}{{\tt 1007.0027}}].

\bibitem{Kobayashi:2010cm}
T.~Kobayashi, M.~Yamaguchi and J.~Yokoyama, \emph{{G-inflation: Inflation
  driven by the Galileon field}},
  \href{http://dx.doi.org/10.1103/PhysRevLett.105.231302}{\emph{Phys. Rev.
  Lett.} {\bf 105} (2010) 231302}, [\href{https://arxiv.org/abs/1008.0603}{{\tt
  1008.0603}}].

\bibitem{Deffayet:2010qz}
C.~Deffayet, O.~Pujolas, I.~Sawicki and A.~Vikman, \emph{{Imperfect Dark Energy
  from Kinetic Gravity Braiding}},
  \href{http://dx.doi.org/10.1088/1475-7516/2010/10/026}{\emph{JCAP} {\bf 10}
  (2010) 026}, [\href{https://arxiv.org/abs/1008.0048}{{\tt 1008.0048}}].

\bibitem{Qiu:2011cy}
T.~Qiu, J.~Evslin, Y.-F. Cai, M.~Li and X.~Zhang, \emph{{Bouncing Galileon
  Cosmologies}},
  \href{http://dx.doi.org/10.1088/1475-7516/2011/10/036}{\emph{JCAP} {\bf 10}
  (2011) 036}, [\href{https://arxiv.org/abs/1108.0593}{{\tt 1108.0593}}].

\bibitem{Easson:2011zy}
D.~A. Easson, I.~Sawicki and A.~Vikman, \emph{{G-Bounce}},
  \href{http://dx.doi.org/10.1088/1475-7516/2011/11/021}{\emph{JCAP} {\bf 11}
  (2011) 021}, [\href{https://arxiv.org/abs/1109.1047}{{\tt 1109.1047}}].

\bibitem{Cai:2012va}
Y.-F. Cai, D.~A. Easson and R.~Brandenberger, \emph{{Towards a Nonsingular
  Bouncing Cosmology}},
  \href{http://dx.doi.org/10.1088/1475-7516/2012/08/020}{\emph{JCAP} {\bf 08}
  (2012) 020}, [\href{https://arxiv.org/abs/1206.2382}{{\tt 1206.2382}}].

\bibitem{Rubakov13-1}
V.~A. Rubakov, \emph{{Consistent NEC-violation: towards creating a universe in
  the laboratory}},
  \href{http://dx.doi.org/10.1103/PhysRevD.88.044015}{\emph{Phys. Rev.} {\bf
  D88} (2013) 044015}, [\href{https://arxiv.org/abs/1305.2614}{{\tt
  1305.2614}}].

\bibitem{Ijjas16-1}
A.~Ijjas and P.~J. Steinhardt, \emph{{Fully stable cosmological solutions with
  a non-singular classical bounce}},
  \href{http://dx.doi.org/10.1016/j.physletb.2016.11.047}{\emph{Phys. Lett.}
  {\bf B764} (2017) 289--294}, [\href{https://arxiv.org/abs/1609.01253}{{\tt
  1609.01253}}].

\bibitem{Monitor:2017mdv}
{\scshape LIGO Scientific, Virgo, Fermi-GBM, INTEGRAL} collaboration, B.~P.
  Abbott et~al., \emph{{Gravitational Waves and Gamma-rays from a Binary
  Neutron Star Merger: GW170817 and GRB 170817A}},
  \href{http://dx.doi.org/10.3847/2041-8213/aa920c}{\emph{Astrophys. J. Lett.}
  {\bf 848} (2017) L13}, [\href{https://arxiv.org/abs/1710.05834}{{\tt
  1710.05834}}].

\bibitem{Kase:2018aps}
R.~Kase and S.~Tsujikawa, \emph{{Dark energy in Horndeski theories after
  GW170817: A review}},
  \href{http://dx.doi.org/10.1142/S0218271819420057}{\emph{Int. J. Mod. Phys.
  D} {\bf 28} (2019) 1942005}, [\href{https://arxiv.org/abs/1809.08735}{{\tt
  1809.08735}}].

\bibitem{Zwiebach85-1}
B.~Zwiebach, \emph{Curvature squared terms in string theories}, {\emph{Phys.
  Lett. B} {\bf 156} (1985) }.

\bibitem{Sen:1985qt}
A.~Sen, \emph{{Equations of Motion for the Heterotic String Theory from the
  Conformal Invariance of the Sigma Model}},
  \href{http://dx.doi.org/10.1103/PhysRevLett.55.1846}{\emph{Phys. Rev. Lett.}
  {\bf 55} (1985) 1846}.

\bibitem{Gross86-1}
D.~J. Gross and E.~Witten, \emph{Superstring modifications of einstein's
  equations}, {\emph{Nucl. Phys. B} {\bf 277} (1986) }.

\bibitem{Gross:1986mw}
D.~J. Gross and J.~H. Sloan, \emph{{The Quartic Effective Action for the
  Heterotic String}},
  \href{http://dx.doi.org/10.1016/0550-3213(87)90465-2}{\emph{Nucl. Phys. B}
  {\bf 291} (1987) 41--89}.

\bibitem{Metsaev:1987zx}
R.~Metsaev and A.~A. Tseytlin, \emph{{Order alpha-prime (Two Loop) Equivalence
  of the String Equations of Motion and the Sigma Model Weyl Invariance
  Conditions: Dependence on the Dilaton and the Antisymmetric Tensor}},
  \href{http://dx.doi.org/10.1016/0550-3213(87)90077-0}{\emph{Nucl. Phys. B}
  {\bf 293} (1987) 385--419}.

\bibitem{VanAcoleyen11-1}
K.~Van~Acoleyen and J.~Van~Doorsselaere, \emph{{Galileons from Lovelock
  actions}}, \href{http://dx.doi.org/10.1103/PhysRevD.83.084025}{\emph{Phys.
  Rev.} {\bf D83} (2011) 084025}, [\href{https://arxiv.org/abs/1102.0487}{{\tt
  1102.0487}}].

\bibitem{Lu:2020iav}
H.~Lu and Y.~Pang, \emph{{Horndeski Gravity as $D\rightarrow4$ Limit of
  Gauss-Bonnet}},  \href{https://arxiv.org/abs/2003.11552}{{\tt 2003.11552}}.

\bibitem{Fernandes:2020nbq}
P.~G. Fernandes, P.~Carrilho, T.~Clifton and D.~J. Mulryne, \emph{{Derivation
  of Regularized Field Equations for the Einstein-Gauss-Bonnet Theory in Four
  Dimensions}},  \href{https://arxiv.org/abs/2004.08362}{{\tt 2004.08362}}.

\bibitem{Hennigar:2020lsl}
R.~A. Hennigar, D.~Kubiznak, R.~B. Mann and C.~Pollack, \emph{{On Taking the
  $D\to 4$ limit of Gauss-Bonnet Gravity: Theory and Solutions}},
  \href{https://arxiv.org/abs/2004.09472}{{\tt 2004.09472}}.

\bibitem{Easson:2020mpq}
D.~A. Easson, T.~Manton and A.~Svesko, \emph{{$D\to4$ Einstein-Gauss-Bonnet
  Gravity and Beyond}},
  \href{http://dx.doi.org/10.1088/1475-7516/2020/10/026}{\emph{JCAP} {\bf 10}
  (2020) 026}, [\href{https://arxiv.org/abs/2005.12292}{{\tt 2005.12292}}].

\bibitem{Easson:2020bgk}
D.~Easson, T.~Manton, M.~Parikh and A.~Svesko, \emph{{The Stringy Origins of
  Galileons and their Novel Limit}},
  \href{http://dx.doi.org/10.1088/1475-7516/2021/05/031}{\emph{JCAP} {\bf 05}
  (2021) 031}, [\href{https://arxiv.org/abs/2012.12277}{{\tt 2012.12277}}].

\bibitem{Montefalcone:2020vlu}
G.~Montefalcone, P.~J. Steinhardt and D.~H. Wesley, \emph{{Dark energy, extra
  dimensions, and the Swampland}},
  \href{http://dx.doi.org/10.1007/JHEP06(2020)091}{\emph{JHEP} {\bf 06} (2020)
  091}, [\href{https://arxiv.org/abs/2005.01143}{{\tt 2005.01143}}].

\bibitem{Gouteraux:2011qh}
B.~Gouteraux, J.~Smolic, M.~Smolic, K.~Skenderis and M.~Taylor,
  \emph{{Holography for Einstein-Maxwell-dilaton theories from generalized
  dimensional reduction}},
  \href{http://dx.doi.org/10.1007/JHEP01(2012)089}{\emph{JHEP} {\bf 01} (2012)
  089}, [\href{https://arxiv.org/abs/1110.2320}{{\tt 1110.2320}}].

\bibitem{Charmousis12-1}
C.~Charmousis, B.~Gouteraux and E.~Kiritsis, \emph{{Higher-derivative
  scalar-vector-tensor theories: black holes, Galileons, singularity cloaking
  and holography}},
  \href{http://dx.doi.org/10.1007/JHEP09(2012)011}{\emph{JHEP} {\bf 09} (2012)
  011}, [\href{https://arxiv.org/abs/1206.1499}{{\tt 1206.1499}}].

\bibitem{Charmousis:2011bf}
C.~Charmousis, E.~J. Copeland, A.~Padilla and P.~M. Saffin, \emph{{General
  second order scalar-tensor theory, self tuning, and the Fab Four}},
  \href{http://dx.doi.org/10.1103/PhysRevLett.108.051101}{\emph{Phys. Rev.
  Lett.} {\bf 108} (2012) 051101}, [\href{https://arxiv.org/abs/1106.2000}{{\tt
  1106.2000}}].

\bibitem{Bakhmatov:2019dow}
I.~Bakhmatov, N.~S. Deger, E.~T. Musaev, E.~O. Colg\'ain and M.~M.
  Sheikh-Jabbari, \emph{{Tri-vector deformations in $d=11$ supergravity}},
  \href{http://dx.doi.org/10.1007/JHEP08(2019)126}{\emph{JHEP} {\bf 08} (2019)
  126}, [\href{https://arxiv.org/abs/1906.09052}{{\tt 1906.09052}}].

\bibitem{DeFelice:2011uc}
A.~De~Felice and S.~Tsujikawa, \emph{{Inflationary non-Gaussianities in the
  most general second-order scalar-tensor theories}},
  \href{http://dx.doi.org/10.1103/PhysRevD.84.083504}{\emph{Phys. Rev. D} {\bf
  84} (2011) 083504}, [\href{https://arxiv.org/abs/1107.3917}{{\tt
  1107.3917}}].

\bibitem{Gasperini07-1}
M.~Gasperini, \emph{{Elements of string cosmology}}.
\newblock Cambridge University Press, 2007.

\bibitem{Maeda:2011zn}
K.-i. Maeda, N.~Ohta and R.~Wakebe, \emph{{Accelerating Universes in String
  Theory via Field Redefinition}},
  \href{http://dx.doi.org/10.1140/epjc/s10052-012-1949-6}{\emph{Eur. Phys. J.
  C} {\bf 72} (2012) 1949}, [\href{https://arxiv.org/abs/1111.3251}{{\tt
  1111.3251}}].

\bibitem{Kanitscheider:2009as}
I.~Kanitscheider and K.~Skenderis, \emph{{Universal hydrodynamics of
  non-conformal branes}},
  \href{http://dx.doi.org/10.1088/1126-6708/2009/04/062}{\emph{JHEP} {\bf 04}
  (2009) 062}, [\href{https://arxiv.org/abs/0901.1487}{{\tt 0901.1487}}].

\bibitem{Chatterjee:2012zh}
S.~Chatterjee, D.~A. Easson and M.~Parikh, \emph{{Energy conditions in the
  Jordan frame}},
  \href{http://dx.doi.org/10.1088/0264-9381/30/23/235031}{\emph{Class. Quant.
  Grav.} {\bf 30} (2013) 235031}, [\href{https://arxiv.org/abs/1212.6430}{{\tt
  1212.6430}}].

\bibitem{Parikh:2015ret}
M.~Parikh and A.~Svesko, \emph{{Thermodynamic Origin of the Null Energy
  Condition}}, \href{http://dx.doi.org/10.1103/PhysRevD.95.104002}{\emph{Phys.
  Rev. D} {\bf 95} (2017) 104002},
  [\href{https://arxiv.org/abs/1511.06460}{{\tt 1511.06460}}].

\bibitem{Parikh:2016lys}
M.~Parikh and A.~Svesko, \emph{{Logarithmic corrections to gravitational
  entropy and the null energy condition}},
  \href{http://dx.doi.org/10.1016/j.physletb.2016.07.071}{\emph{Phys. Lett. B}
  {\bf 761} (2016) 16--19}, [\href{https://arxiv.org/abs/1612.06949}{{\tt
  1612.06949}}].

\bibitem{Veneziano:2000pz}
G.~Veneziano, \emph{{String cosmology: The Pre - big bang scenario}},  in
  \emph{{71st Les Houches Summer School: The Primordial Universe}},
  pp.~581--628, 2000.
\newblock \href{https://arxiv.org/abs/hep-th/0002094}{{\tt hep-th/0002094}}.
\newblock \href{http://dx.doi.org/10.1007/3-540-45334-2_12}{DOI}.

\bibitem{Gasperini:2002bn}
M.~Gasperini and G.~Veneziano, \emph{{The Pre - big bang scenario in string
  cosmology}},
  \href{http://dx.doi.org/10.1016/S0370-1573(02)00389-7}{\emph{Phys. Rept.}
  {\bf 373} (2003) 1--212}, [\href{https://arxiv.org/abs/hep-th/0207130}{{\tt
  hep-th/0207130}}].

\bibitem{Gasperini:1992em}
M.~Gasperini and G.~Veneziano, \emph{{Pre - big bang in string cosmology}},
  \href{http://dx.doi.org/10.1016/0927-6505(93)90017-8}{\emph{Astropart. Phys.}
  {\bf 1} (1993) 317--339}, [\href{https://arxiv.org/abs/hep-th/9211021}{{\tt
  hep-th/9211021}}].

\bibitem{Gasperini:1991ak}
M.~Gasperini and G.~Veneziano, \emph{{O(d,d) covariant string cosmology}},
  \href{http://dx.doi.org/10.1016/0370-2693(92)90744-O}{\emph{Phys. Lett. B}
  {\bf 277} (1992) 256--264}, [\href{https://arxiv.org/abs/hep-th/9112044}{{\tt
  hep-th/9112044}}].

\bibitem{Gasperini:1996np}
M.~Gasperini, J.~Maharana and G.~Veneziano, \emph{{Graceful exit in quantum
  string cosmology}},
  \href{http://dx.doi.org/10.1016/0550-3213(96)00201-5}{\emph{Nucl. Phys. B}
  {\bf 472} (1996) 349--360}, [\href{https://arxiv.org/abs/hep-th/9602087}{{\tt
  hep-th/9602087}}].

\bibitem{Gasperini:1996fu}
M.~Gasperini, M.~Maggiore and G.~Veneziano, \emph{{Towards a nonsingular pre -
  big bang cosmology}},
  \href{http://dx.doi.org/10.1016/S0550-3213(97)00149-1}{\emph{Nucl. Phys. B}
  {\bf 494} (1997) 315--330}, [\href{https://arxiv.org/abs/hep-th/9611039}{{\tt
  hep-th/9611039}}].

\bibitem{Banks:2004cw}
T.~Banks, W.~Fischler and L.~Mannelli, \emph{{Microscopic quantum mechanics of
  the p = rho universe}},
  \href{http://dx.doi.org/10.1103/PhysRevD.71.123514}{\emph{Phys. Rev. D} {\bf
  71} (2005) 123514}, [\href{https://arxiv.org/abs/hep-th/0408076}{{\tt
  hep-th/0408076}}].

\bibitem{Banks:2004vg}
T.~Banks and W.~Fischler, \emph{{Holographic cosmology}},
  \href{https://arxiv.org/abs/hep-th/0405200}{{\tt hep-th/0405200}}.

\bibitem{Banks:2003ta}
T.~Banks and W.~Fischler, \emph{{Holographic cosmology 3.0}},
  \href{http://dx.doi.org/10.1238/Physica.Topical.117a00056}{\emph{Phys.
  Scripta T} {\bf 117} (2005) 56--63},
  [\href{https://arxiv.org/abs/hep-th/0310288}{{\tt hep-th/0310288}}].

\bibitem{Battefeld:2004jh}
T.~J. Battefeld and D.~A. Easson, \emph{{Perturbations in a holographic
  universe and in other stiff fluid cosmologies}},
  \href{http://dx.doi.org/10.1103/PhysRevD.70.103516}{\emph{Phys. Rev. D} {\bf
  70} (2004) 103516}, [\href{https://arxiv.org/abs/hep-th/0408154}{{\tt
  hep-th/0408154}}].

\bibitem{Bousso:1999cb}
R.~Bousso, \emph{{Holography in general space-times}},
  \href{http://dx.doi.org/10.1088/1126-6708/1999/06/028}{\emph{JHEP} {\bf 06}
  (1999) 028}, [\href{https://arxiv.org/abs/hep-th/9906022}{{\tt
  hep-th/9906022}}].

\bibitem{DeFelice:2011bh}
A.~De~Felice and S.~Tsujikawa, \emph{{Conditions for the cosmological viability
  of the most general scalar-tensor theories and their applications to extended
  Galileon dark energy models}},
  \href{http://dx.doi.org/10.1088/1475-7516/2012/02/007}{\emph{JCAP} {\bf 02}
  (2012) 007}, [\href{https://arxiv.org/abs/1110.3878}{{\tt 1110.3878}}].

\bibitem{Adams:2006sv}
A.~Adams, N.~Arkani-Hamed, S.~Dubovsky, A.~Nicolis and R.~Rattazzi,
  \emph{{Causality, analyticity and an IR obstruction to UV completion}},
  \href{http://dx.doi.org/10.1088/1126-6708/2006/10/014}{\emph{JHEP} {\bf 10}
  (2006) 014}, [\href{https://arxiv.org/abs/hep-th/0602178}{{\tt
  hep-th/0602178}}].

\bibitem{Shore:2007um}
G.~M. Shore, \emph{{Superluminality and UV completion}},
  \href{http://dx.doi.org/10.1016/j.nuclphysb.2007.03.034}{\emph{Nucl. Phys. B}
  {\bf 778} (2007) 219--258}, [\href{https://arxiv.org/abs/hep-th/0701185}{{\tt
  hep-th/0701185}}].

\bibitem{Dobre:2017pnt}
D.~A. Dobre, A.~V. Frolov, J.~T. G\'alvez~Ghersi, S.~Ramazanov and A.~Vikman,
  \emph{{Unbraiding the Bounce: Superluminality around the Corner}},
  \href{http://dx.doi.org/10.1088/1475-7516/2018/03/020}{\emph{JCAP} {\bf 03}
  (2018) 020}, [\href{https://arxiv.org/abs/1712.10272}{{\tt 1712.10272}}].

\bibitem{deRham:2021fpu}
C.~de~Rham, S.~Melville and J.~Noller, \emph{{Positivity Bounds on Dark Energy:
  When Matter Matters}},  \href{https://arxiv.org/abs/2103.06855}{{\tt
  2103.06855}}.

\bibitem{Easson:2013bda}
D.~A. Easson, I.~Sawicki and A.~Vikman, \emph{{When Matter Matters}},
  \href{http://dx.doi.org/10.1088/1475-7516/2013/07/014}{\emph{JCAP} {\bf 07}
  (2013) 014}, [\href{https://arxiv.org/abs/1304.3903}{{\tt 1304.3903}}].

\bibitem{Bains:2015gpv}
J.~S. Bains, M.~P. Hertzberg and F.~Wilczek, \emph{{Oscillatory Attractors: A
  New Cosmological Phase}},
  \href{http://dx.doi.org/10.1088/1475-7516/2017/05/011}{\emph{JCAP} {\bf 05}
  (2017) 011}, [\href{https://arxiv.org/abs/1512.02304}{{\tt 1512.02304}}].

\bibitem{Easson:2016klq}
D.~A. Easson and A.~Vikman, \emph{{The Phantom of the New Oscillatory
  Cosmological Phase}},  \href{https://arxiv.org/abs/1607.00996}{{\tt
  1607.00996}}.

\bibitem{Easson:2018qgr}
D.~A. Easson and T.~Manton, \emph{{Stable Cosmic Time Crystals}},
  \href{http://dx.doi.org/10.1103/PhysRevD.99.043507}{\emph{Phys. Rev. D} {\bf
  99} (2019) 043507}, [\href{https://arxiv.org/abs/1802.03693}{{\tt
  1802.03693}}].

\bibitem{Padilla:2010de}
A.~Padilla, P.~M. Saffin and S.-Y. Zhou, \emph{{Bi-galileon theory I:
  Motivation and formulation}},
  \href{http://dx.doi.org/10.1007/JHEP12(2010)031}{\emph{JHEP} {\bf 12} (2010)
  031}, [\href{https://arxiv.org/abs/1007.5424}{{\tt 1007.5424}}].

\bibitem{Padilla:2010ir}
A.~Padilla, P.~M. Saffin and S.-Y. Zhou, \emph{{Multi-galileons, solitons and
  Derrick's theorem}},
  \href{http://dx.doi.org/10.1103/PhysRevD.83.045009}{\emph{Phys. Rev. D} {\bf
  83} (2011) 045009}, [\href{https://arxiv.org/abs/1008.0745}{{\tt
  1008.0745}}].

\bibitem{Deffayet:2010zh}
C.~Deffayet, S.~Deser and G.~Esposito-Farese, \emph{{Arbitrary $p$-form
  Galileons}}, \href{http://dx.doi.org/10.1103/PhysRevD.82.061501}{\emph{Phys.
  Rev. D} {\bf 82} (2010) 061501}, [\href{https://arxiv.org/abs/1007.5278}{{\tt
  1007.5278}}].

\bibitem{Hinterbichler:2010xn}
K.~Hinterbichler, M.~Trodden and D.~Wesley, \emph{{Multi-field galileons and
  higher co-dimension branes}},
  \href{http://dx.doi.org/10.1103/PhysRevD.82.124018}{\emph{Phys. Rev. D} {\bf
  82} (2010) 124018}, [\href{https://arxiv.org/abs/1008.1305}{{\tt
  1008.1305}}].

\bibitem{Vafa:2005ui}
C.~Vafa, \emph{{The String landscape and the swampland}},
  \href{https://arxiv.org/abs/hep-th/0509212}{{\tt hep-th/0509212}}.

\bibitem{Ooguri:2006in}
H.~Ooguri and C.~Vafa, \emph{{On the Geometry of the String Landscape and the
  Swampland}},
  \href{http://dx.doi.org/10.1016/j.nuclphysb.2006.10.033}{\emph{Nucl. Phys. B}
  {\bf 766} (2007) 21--33}, [\href{https://arxiv.org/abs/hep-th/0605264}{{\tt
  hep-th/0605264}}].

\bibitem{Palti:2019pca}
E.~Palti, \emph{{The Swampland: Introduction and Review}},
  \href{http://dx.doi.org/10.1002/prop.201900037}{\emph{Fortsch. Phys.} {\bf
  67} (2019) 1900037}, [\href{https://arxiv.org/abs/1903.06239}{{\tt
  1903.06239}}].

\bibitem{vanBeest:2021lhn}
M.~van Beest, J.~Calder\'on-Infante, D.~Mirfendereski and I.~Valenzuela,
  \emph{{Lectures on the Swampland Program in String Compactifications}},
  \href{https://arxiv.org/abs/2102.01111}{{\tt 2102.01111}}.

\bibitem{Heisenberg:2019qxz}
L.~Heisenberg, M.~Bartelmann, R.~Brandenberger and A.~Refregier,
  \emph{{Horndeski gravity in the swampland}},
  \href{http://dx.doi.org/10.1103/PhysRevD.99.124020}{\emph{Phys. Rev. D} {\bf
  99} (2019) 124020}, [\href{https://arxiv.org/abs/1902.03939}{{\tt
  1902.03939}}].

\bibitem{Brahma:2019kch}
S.~Brahma and M.~W. Hossain, \emph{{Dark energy beyond quintessence:
  Constraints from the swampland}},
  \href{http://dx.doi.org/10.1007/JHEP06(2019)070}{\emph{JHEP} {\bf 06} (2019)
  070}, [\href{https://arxiv.org/abs/1902.11014}{{\tt 1902.11014}}].

\bibitem{Agrawal:2018own}
P.~Agrawal, G.~Obied, P.~J. Steinhardt and C.~Vafa, \emph{{On the Cosmological
  Implications of the String Swampland}},
  \href{http://dx.doi.org/10.1016/j.physletb.2018.07.040}{\emph{Phys. Lett. B}
  {\bf 784} (2018) 271--276}, [\href{https://arxiv.org/abs/1806.09718}{{\tt
  1806.09718}}].

\bibitem{Obied:2018sgi}
G.~Obied, H.~Ooguri, L.~Spodyneiko and C.~Vafa, \emph{{De Sitter Space and the
  Swampland}},  \href{https://arxiv.org/abs/1806.08362}{{\tt 1806.08362}}.

\bibitem{Metsaev87-1}
R.~R. Metsaev and A.~A. Tseytlin, \emph{{Order alpha-prime (Two Loop)
  Equivalence of the String Equations of Motion and the Sigma Model Weyl
  Invariance Conditions: Dependence on the Dilaton and the Antisymmetric
  Tensor}}, \href{http://dx.doi.org/10.1016/0550-3213(87)90077-0}{\emph{Nucl.
  Phys.} {\bf B293} (1987) 385--419}.

\bibitem{Zwiebach:1985uq}
B.~Zwiebach, \emph{{Curvature Squared Terms and String Theories}},
  \href{http://dx.doi.org/10.1016/0370-2693(85)91616-8}{\emph{Phys. Lett. B}
  {\bf 156} (1985) 315--317}.

\bibitem{Kobayashi:2019hrl}
T.~Kobayashi, \emph{{Horndeski theory and beyond: a review}},
  \href{http://dx.doi.org/10.1088/1361-6633/ab2429}{\emph{Rept. Prog. Phys.}
  {\bf 82} (2019) 086901}, [\href{https://arxiv.org/abs/1901.07183}{{\tt
  1901.07183}}].

\end{thebibliography}\endgroup
\end{document}